\documentclass[aps,pre,twoside,12pt,nofootinbib,longbibliography]{revtex4-1}
\usepackage{amsmath,amsxtra,amsopn,amsbsy,amstext,amsfonts,amsthm,amssymb}
\usepackage[colorlinks=true, citecolor=red, urlcolor=blue ]{hyperref}
\usepackage{amsmath,amsxtra,amsopn,amsbsy,amstext,amsfonts,amsthm,amssymb,array,bbm,dsfont,xcolor,comment,graphicx,graphics}
\usepackage[normalem]{ulem}
\usepackage[tmargin=3cm,bmargin=3cm,lmargin=3cm,rmargin=3cm]{geometry}
\usepackage[mathscr]{euscript}

\begin{document}

\title{Exactly Solvable 1D Quantum Models with Gamma Matrices}
\author{Yash Chugh, Kusum Dhochak, Uma Divakaran,  Prithvi Narayan, Amit Kumar Pal} 
\affiliation{Department of Physics, Indian Institute of Technology Palakkad, Palakkad 678 557, India} 
\date{\today}

\begin{abstract}
In this paper, we write exactly solvable generalizations of $1$-dimensional quantum XY and Ising-like models by using $2^d$-dimensional {Gamma ($\Gamma$)} matrices as the degrees of freedom on each site. We show that these models result in quadratic Fermionic Hamiltonians with Jordan-Wigner like transformations.  We illustrate the techniques using a specific case of $4$-dimensional $\Gamma$ matrices and explore the quantum phase transitions present in the model.
\end{abstract}

\maketitle 

\section{Introduction}
\label{sec:introduction}

Investigations of exactly solvable quantum many-body models are important due to their immense applications in understanding a plethora of physical phenomena of interests in statistical and condensed-matter physics, such as quantum phase transitions~\cite{Sachdev2011,Dutta2015} and thermodynamic properties of many-body systems~\cite{Takahashi1999}. They not only provide platforms for testing out new approximation schemes, but also serve as test-beds for numerical techniques developed to tackle large many-body systems, in particular, higher dimensional systems or non-integrable systems ~\cite{Schollwock2011,Pang2016,Montanegro2018,cirac_mps_2006,cirac_mps_2008}.

Despite their enormous importance in various fields, only a handful of exactly solvable quantum many-body models are known till date, mostly in one dimension~\cite{Zvyagin2010,doi:10.1142/p364,Giamarchi:743140,Franchini:2016cxs,doi:10.1080/00107514.2016.1259251}. Among such models, perhaps the most celebrated ones are the Ising model~\cite{Elliot1970,Pfeuty1970,Elliott1971,*Pfeuty1971,Stinchcombe1973,*Stinchcombe1973a,*Stinchcombe1973b} and the XY model~\cite{Lieb1961,Katsura1962,Barouch1970,*Barouch1971,*Barouch1971a} in a transverse field (see also~\cite{Kogut1979}), consisting of a number of spin-$\frac{1}{2}$ particles arranged on a one-dimensional lattice. These models have a rich history of aiding research in various directions over the years, including understanding order-disorder quantum phase transitions~\cite{Sachdev2011,Dutta2015}, quantum information science and technology~\cite{Amico2008} and material research in condensed matter physics~\cite{aeppli_82}.  Moreover, realization of these models through currently available techniques using different substrates such as trapped ions~\cite{Porras2004,*Deng2005}, nuclear magnetic resonance systems~\cite{Zhang2012}, solid-state systems~\cite{Schechter2008}, and optical lattices~\cite{Duan2003,*Simon2011,*Liao2021} have made the verification of theoretical results possible. 

While the simplest variants of the Ising and the XY models deal with the spin-$\frac{1}{2}$ particles arranged on a one-dimensional lattice where a spin only interacts with its nearest-neighbours, these models have been further extended in various directions. For example, one could have asymmetric Dzyaloshinskii-Moriya-type interactions~\cite{SISKENS1975259}, with staggered magnetic field~\cite{divakaran2008}, and multiple spin-exchange interactions~\cite{Kopp2005,Zvyagin2006,Zvyagin2009}. The Ising and the XY model in a transverse field, along with its generalizations mentioned above, can be solved by transforming the spin variables to spinless fermions via a Jordan-Wigner (JW) transformation~\cite{Wigner1928},
followed by a Bogoliubov-de Gennes transformation. 

In this paper, we explore one such {exactly solvable} generalization
%, where it is extended to a model 
with higher-dimensional Hilbert space associated to  each lattice site. Noticing that the anti-commutation relations of Pauli operators, i.e., $\{ \sigma^i,\sigma^j \}= 2 \delta^{ij}$, $i,j \in (1,2,3)$, play a crucial role in the JW approach, {earlier works \cite{1998NuPhB.535..681D} have proposed replacing Pauli matrices by} higher-dimensional Gamma matrices, $\Gamma^i, i \in (1,2,3,\dots)$, satisfying similar anti-commutation relations, i.e., 
\begin{eqnarray}
\{ \sigma^i,\sigma^j \}= 2 \delta^{ij} \quad  \longrightarrow \quad  \{ \Gamma^i ,\Gamma^j \} = 2 \delta^{ij}.
\end{eqnarray}
It was shown in  \cite{1998NuPhB.535..681D} that  such a model can be fermionized just like XY models. In this work, we construct explicitly models with $2^d-$dimensional Gamma matrices (for any $d$), which when fermionized, give Hamiltonians which are quadratic in fermions and hence solvable. We also discuss the $d=2$ case in more detail and explore an Ising-like quantum critical point.

The rest of the  paper is organized as follows : In section \ref{sec:model}, after reviewing the one-dimensional solvable XY and related models, we define our model using the Gamma matrices. We then rewrite the model in terms of fermions, employing the JW transformation, and solve it. In section \ref{sec:special_case}, we illustrate our results via solving the model explicitly for a special case. We also demonstrate the quantum phase transitions occurring in the system, and comment on the calculation of critical exponents. We conclude in section \ref{sec:results}, pointing out possible future directions.  

\section{Model}
\label{sec:model}

Let us begin by quickly recalling a class of quantum spin models consisting of a lattice of $N$ sites with two degrees of freedom at each site. The Hamiltonian representing such models can be written in a compact form as
\begin{equation}
    H  =- i  \sum_{a} \sum_{\mu,\nu=1}^{2} J_{\mu \nu} \sigma^{\mu}_{a} \sigma^3_a \sigma^{\nu}_{a+1}  - h \sum_a \sigma^3_a,
    \label{eq:Ising_Hamiltonian}
\end{equation} 
where $\sigma^1,\sigma^2$ and $\sigma^3$ are the Pauli matrices, and $a= 1,\dots N$ is the lattice index, such that on each site
\begin{equation}
    \{ \sigma^\mu , \sigma^\nu \} = 2 \delta^{\mu \nu}.
\end{equation}
The spin-exchange couplings are represented by $J_{\mu\nu}$, while the strengths of the external magnetic field along the $z$-direction is denoted by $h$. It is worthwhile to mention that the above Hamiltonian can be easily re-written in a more familiar form that is quadratic in Pauli matrices using $\sigma^1 \sigma^2 = i \sigma^3$ and so on. However we will work with (\ref{eq:Ising_Hamiltonian}), since it is better suited to the generalizations that we will define later.

A number of well-known quantum spin models with nearest-neighbor spin-exchange interactions can be identified as particular cases of the Hamiltonian in (\ref{eq:Ising_Hamiltonian}), as follows. \begin{enumerate}
    \item Ising model in a transverse field :  $J_{21} =  J_1,$  rest of $J_{\mu \nu}$ vanishing 
    \begin{eqnarray}
    H &=& \sum_{a} \left(  J_1 \sigma^{1}_{a}  \sigma^{1}_{a+1}  - h  \sigma^3_a \right)
    \label{eq:tfim}
    \end{eqnarray}
    \item XY model in a transverse field: $J_{12} = -J_2, J_{21} = J_1,$ rest of $J_{\mu \nu}$ vanishing
    \begin{eqnarray}
    H &=& \sum_{a} \left(  J_1 \sigma^{1}_{a}  \sigma^{1}_{a+1} + J_2 \sigma^{2}_{a}  \sigma^{2}_{a+1}  - h  \sigma^3_a \right)
    \label{eq:xy}
    \end{eqnarray}    
    \item XY model in a transverse field with asymmetric Dzyaloshinskii–Moriya (DM) interaction:  $J_{12} =- J_2, J_{21} =  J_1, J_{11}=J_{22} = D,$ rest of $J_{\mu \nu}$ vanishing
    \begin{eqnarray}
    H &=&\sum_{a} \left(  J_1 \sigma^{1}_{a}  \sigma^{1}_{a+1} + J_2 \sigma^{2}_{a}\sigma^{2}_{a+1}  + D (  \sigma^{1}_{a}  \sigma^{2}_{a+1} -  \sigma^{2}_{a}  \sigma^{1}_{a+1})   - h  \sigma^3_a \right)
    \label{eq:xydm}
    \end{eqnarray}    
\end{enumerate}
For convenience, we shall refer to the class of quantum spin models represented by (\ref{eq:Ising_Hamiltonian}) as the  \emph{generalized-XY} (g-XY) models.  As mentioned earlier, these models are solvable using the JW transformations, which rewrite Pauli matrices in terms of fermionic creation and annihilation operators obeying canonical anticommutation relations, as we will see below. The Hamiltonian (\ref{eq:Ising_Hamiltonian}) is quadratic in terms of these fermionic operators, and hence solvable.

In what follows, we generalize the g-XY model to allow for more degrees of freedom per lattice site in such a way that the JW transformations remain applicable, and the resulting Hamiltonian remains quadratic in terms of the fermionic operators. As noted in \cite{1998NuPhB.535..681D}, such a generalization is possible by replacing the Pauli matrices on each lattice site with appropriate $\Gamma$-matrices\footnote{ See \cite{2021arXiv210706335B} and references therein for a recent exposition on the general conditions when such rewriting is possible} , which have the following algebra at each site:
\begin{align} \label{Clifford Algebra}
    \{  \Gamma_a^\mu ,  \Gamma_a^\nu \} = 2 \delta^{\mu \nu},
\end{align}
where $\mu,\nu \in \{1,2.....2d\}$ while $\Gamma-$matrices at different sites commute. For the specific representation of the $\Gamma$-matrices in terms of Pauli matrices see section \ref{subsubsec:Pauli_represetation} - the Pauli matrices $\sigma^\mu_a$ on the lattice site $a$ correspond to the special case of $d=1$. In the next few subsections, we work this generalization out in detail, and demonstrate the solvability of the generalized model (see (\ref{eq:Hamiltonian_in_Fermions}) for the Hamiltonian of the model). We will call this class of models as \emph{generalised XY model with Gamma matrices} (g-XYG), parametrized (apart from the different interaction parameters appearing in the Hamiltonian) by the parameter $d$.

\subsection{Review  of Gamma Matrices}
\label{subsec:gamma_matrices_review}

We begin with reviewing a number of features of the Gamma matrices which will be important in the rest of the paper. For brevity, we suppress the lattice index, and write the anticommutation relation of the Gamma matrices as  
\begin{equation}
\label{eq:Clifford_Algebra_Partial}
    \{ \Gamma^\mu , \Gamma^\nu \} =2 \delta^{\mu \nu},  
\end{equation}
where $\mu=\{1,2,\cdots,2d\}$. As we will see later, (section \ref{subsubsec:Pauli_represetation}) these $\Gamma^\mu$ are $2^{d} \times 2^{d}$ matrices, and can be thought of as operators  acting on a Hilbert space of $d$ spin-$1/2$ degrees of freedom. For $d=1$, it is clear that (\ref{eq:Clifford_Algebra_Partial})  simply reduces to Pauli matrices $\sigma^1,\sigma^2$. The matrix $\Gamma^{2d+1}$, which is the analogue of the $\sigma^3$ matrix for $d=1$ case and plays an important role in defining the Hamiltonian for the g-XYG models (see section \ref{subsec:hamiltonian}), is defined as  
\begin{equation}
    \Gamma^{2d+1} \equiv (-\text{i})^d\prod_{\mu=1}^{2d}\Gamma^{\mu},  
\end{equation}
and obeys the anticommutation relation 
\begin{eqnarray}
    \{\Gamma^{2d+1},\Gamma^{\mu}\} = 0\  \forall \mu,  
\end{eqnarray}
and $(\Gamma^{2d+1})^2 = 1$.   Additionally, we define a set of $d$ mutually commuting operators $S_i$, such that $[S_i,S_j]=0,$ where
\begin{equation}
    S_i \equiv (-\text{i}) \Gamma^{2i-1} \Gamma^{2i} , i \in \{1,2,\cdots,d\}.
\end{equation}
These operators will facilitate the \emph{field term} in the g-XYG model (see section \ref{subsec:hamiltonian}). 

\subsubsection{A specific representation of the \texorpdfstring{$\Gamma$}{Gamma} matrices}
\label{subsubsec:Pauli_represetation} 

\begin{figure*}
    \centering
    \includegraphics[width=0.6\textwidth]{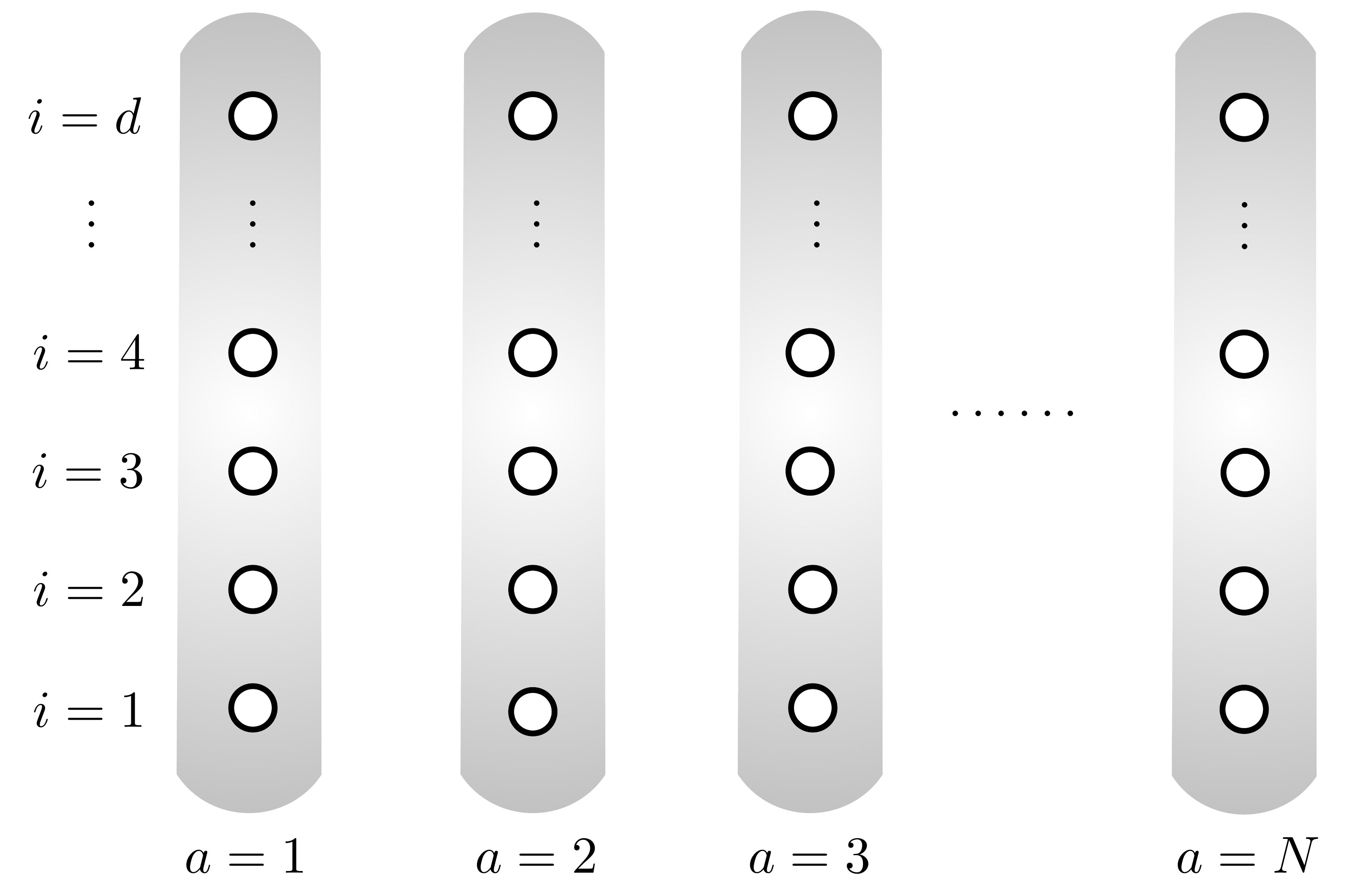}
    \caption{\textbf{Sublattice structure for g-XYG models.} Each lattice site (gray blocks) denoted by the index $a$ consists of $d$ sublattice points (white circles), marked by the index $i$.}
    \label{fig:sublattice}
\end{figure*}

While defining and solving our model can be done purely algebraically (i.e using the algebra defined in (\ref{Clifford Algebra})), it is sometimes useful to have explicit realization for the $\Gamma^\mu_a$ (for each lattice site $a$) operators as $2^{d} \times 2^{d}$ matrices on the Hilbert space. One such realization of the $\Gamma$ matrices is in terms of the tensor products of Pauli Matrices given as (again suppressing the lattice index $a$)
\begin{eqnarray}
\label{eq:Gamma_with_Pauli}
\Gamma^1 &=& \sigma^1_1 \otimes \mathds{1}_2 \otimes \mathds{1}_3 \otimes \cdots \otimes \mathds{1}_{d-1}\otimes\mathds{1}_{d},\nonumber\\
\Gamma^2 &=& \sigma^2_1 \otimes \mathds{1}_2 \otimes \mathds{1}_3 \otimes \cdots \otimes \mathds{1}_{d-1}\otimes\mathds{1}_d \nonumber\\
\Gamma^{3} &=& \sigma^{3}_1 \otimes \sigma^{1}_2\otimes \mathds{1}_3 \otimes \cdots \otimes \mathds{1}_{d-1}\otimes\mathds{1}_d\nonumber\\
\Gamma^{4} &=& \sigma^{3}_1 \otimes \sigma^{2}_2 \otimes \mathds{1}_3 \otimes \cdots \otimes \mathds{1}_{d-1}\otimes \mathds{1}_d\nonumber\\
    &\vdots& \nonumber\\
\Gamma^{2d-1} &=& \sigma^{3}_1 \otimes \sigma^{3}_2 \otimes \sigma^{3}_3  \otimes \cdots\otimes \sigma^3_{d-1}\otimes \sigma^{1}_d \nonumber\\
\Gamma^{2d} &=& \sigma^{3}_1 \otimes \sigma^{3}_2 \otimes \sigma^{3}_3  \otimes \cdots\otimes \sigma^3_{d-1}\otimes \sigma^{2}_d,
\end{eqnarray}
where it can be easily verified that the above matrices satisfy (\ref{eq:Clifford_Algebra_Partial}). One can interpret it as $d$ sublattice sites  for each of the lattice sites $a$, having a spin-$1/2$ degree of freedom on each of these sublattice sites or as $d$ spin-$1/2$ pseudo-spin degrees of freedom at each lattice site. The subscripts on the Pauli matrices and the identity operators in (\ref{eq:Gamma_with_Pauli}) represents the sublattice points/pseudo-spin, which we denote by the index $i$ ($i=1,2,\cdots,d$), as mentioned before (see section~\ref{subsec:gamma_matrices_review}). A pictorial representation of the sublattice structure of the model can be found in Fig.~\ref{fig:sublattice}.

The above representation corresponds to the choice 
\begin{eqnarray}
\Gamma^{2d+1} &=& \bigotimes _{i=1}^d\sigma^3_i=\prod_{i=1}^d S_i,
\end{eqnarray}
where the commuting operators $S_i$ (see section~\ref{subsec:gamma_matrices_review}) can be constructed as
\begin{eqnarray}
S_i={\mathds 1}_1\otimes\mathds{1}_2\otimes\cdots\mathds{1}_{i-1}\otimes\sigma^3_i  \otimes\mathds{1}_{i+1}\otimes\cdots\otimes\mathds{1}_d. 
\end{eqnarray}

\subsection{Jordan-Wigner Transformation}
\label{subsec:JW_transformation}

\begin{figure*}
    \centering
    \includegraphics[width=0.6\textwidth]{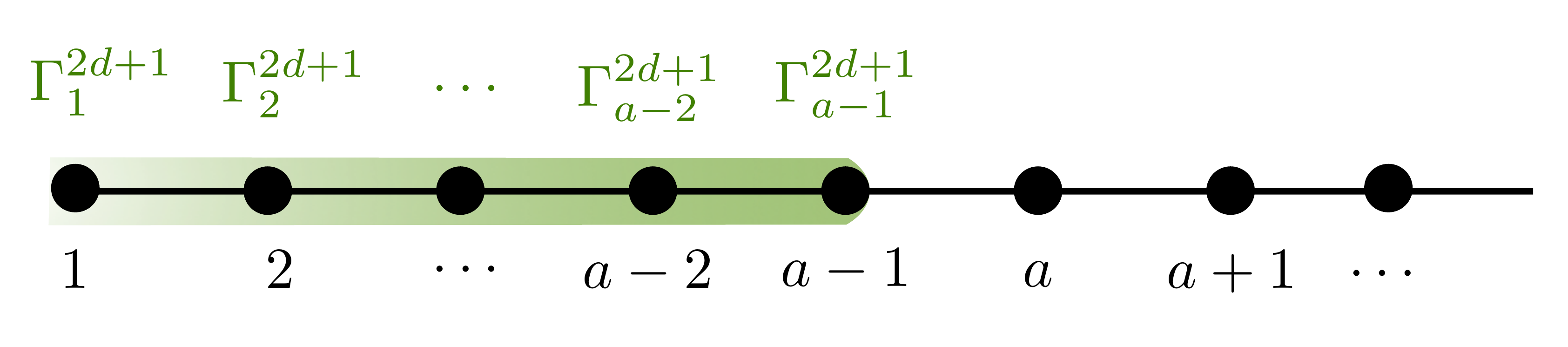}
    \caption{\textbf{Jordan string.} A Jordan string of $\Gamma^{2d+1}$ operators corresponding to the fermionic operator $\chi^\mu_a$, spanning on the lattice sites $1,2,\cdots,a-1$.}
    \label{fig:jordan_string}
\end{figure*}

The model we build below consists of $\Gamma^\mu_a$, i.e Gamma matrices defined at each site $a$. We define the fermion operator $\chi^\mu_a$ as
\begin{equation}
\label{eq:Jordan_Wigner_Fermions_from_spins}
\chi^\mu_a  \equiv \left( \prod_{b < a} \Gamma^{2d+1}_b\right) \Gamma^\mu_a,
\end{equation}
such that ${\chi^\mu_a}^\dagger = \chi^\mu_a$. It is easy to verify (also see \cite{1998NuPhB.535..681D}) that the operators $\{\chi^\mu_a \}$ satisfy the fermionic algebra 
\begin{equation}
   \{ \chi^\mu_a ,\chi^\nu_b\} = 2\delta^{\mu \nu} \delta_{ab},
\end{equation} 
and are called Majorana fermions. Note that this is a straightforward $\Gamma$-matrix generalization of the well  known JW transformations usually defined in terms of Pauli matrices, with the $\Gamma^{2d+1}$ playing the role of $\sigma^3$. The so called \emph{Jordan string} now consists of a string of $\Gamma^{2d+1}$ operators to the left of the site of interest (see Fig.~\ref{fig:jordan_string}). Although the $\Gamma-$matrices at different sites commute, the presence of $\Gamma^{2d+1}$ in the Jordan string of fermions along with the property $\{ \Gamma^\mu_a , \Gamma^{2d+1}_a \} = 0$ makes the fermions at different sites anticommute with each other. An analogous treatment using complex fermions can also be done and some details are given in the appendix \ref{sec:complex_fermions}.

\subsection{Hamiltonian}
\label{subsec:hamiltonian}

\begin{figure*}
    \centering
    \includegraphics[width=0.6\textwidth]{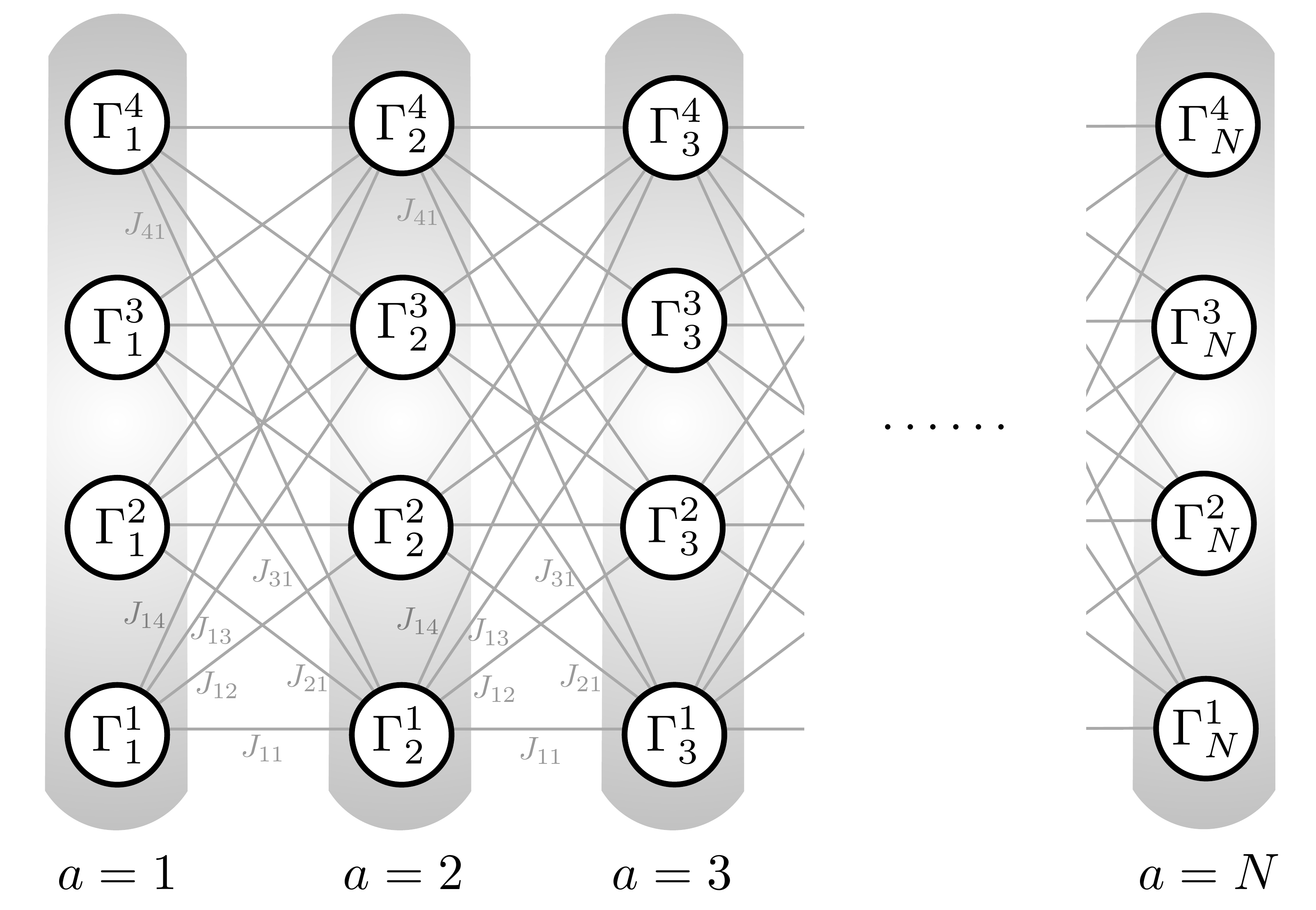}
    \caption{\textbf{Interactions in the Hamiltonian for $d=2$.} Each lattice site $a$ (gray blocks) can house four $\Gamma$ matrices, $\{\Gamma^1_a,\Gamma^2_a,\Gamma^3_a,\Gamma^4_a\}$, $a=1,2,\cdots,N$. The lines in the diagram represents the couplings $J_{\mu\nu}$ involving $\Gamma^\mu_a$ and $\Gamma^{\nu}_{a+1}$, $\mu,\nu\in\{1,2,3,4\}$ (see (\ref{eq:Hamiltonian_in_Gamma})). }
    \label{fig:couplings}
\end{figure*}

With all the ingredients in place, we now write down the g-XYG models as a generalization of the g-XY models in terms of $\Gamma$ matrices. Let us consider the following Hamiltonian: 
\begin{eqnarray}
\label{eq:Hamiltonian_in_Gamma}
    H_G = - \text{i}\sum_{a}\sum_{\mu,\nu}^{} \text{J}_{\mu\nu} \Gamma_{a}^{\mu} \Gamma_{a}^{2d+1} \Gamma_{a+1}^{\nu} - \sum_{a}\sum_{i} h_i S^{i}_{a},  
\end{eqnarray}
where $\mu,\nu = 1,2\;...\;2d$, $i = 1,2\;...\;d$  and $\text{J}_{\mu\nu}$ ($h_i$) are a set of $4d^2$ ($d$) coupling constants, respectively\footnote{As we will show in Appendix \ref{sec:Appedix:More general Hamiltonian}, the number of independent couplings can be shown to be $4d^2$ rather than $4d^2 +d$}. The hermiticity condition of the Hamiltonian implies that the coupling constants $J_{\mu \nu}$ and $h_i$ must be real. Note that since for $d=1$, $\Gamma$ matrices  reduce to Pauli $\sigma$ matrices, the g-XYG Hamiltonian given above reduces to the g-XY Hamiltonian (\ref{eq:Ising_Hamiltonian}) for $d=1$. A pictorial representation of the $J_{\mu \nu}$ couplings for the $d=2$ case can be found in Fig.~\ref{fig:couplings}. We sometimes write the coupling constants $J_{\mu \nu}$ as a sum of a symmetric and an antisymmetric part, as $J_{\mu \nu} ={ (\mathscr{S}_{\mu \nu } - \mathscr{A}_{\mu \nu})/2 }$, where we take $\mathscr{S}_{\mu \nu} = \mathscr{S}_{\nu \mu}$ and $ \mathscr{A}_{\mu \nu} = - \mathscr{A}_{\nu \mu}$. 

As described in Appendix \ref{sec:Finite N Hamiltonian}, the g-XYG models given by the Hamiltonian in (\ref{eq:Hamiltonian_in_Gamma}) are quadratic in terms of the fermions defined by the JW transformations given in (\ref{eq:Jordan_Wigner_Fermions_from_spins})
 \begin{eqnarray}
 \label{eq:Hamiltonian_in_Fermions}
    H_G &=& -\text{i}\sum_{a}\sum_{\mu,\nu}\ \text{J}_{\mu\nu} \chi^\mu_a \chi^\nu_{a+1} + \text{i}\sum_{a}\sum_{i}  h_i \chi^{2i-1}_a \chi^{2i}_a.
\end{eqnarray}
Such quadratic fermionic Hamiltonians are often encountered in literature~\cite{bdg2016}. In the next subsection, we diagonalize the Hamiltonian by exploiting the translational symmetry. In the case of systems with finite $N$, the transformation of the Hamiltonian (\ref{eq:Hamiltonian_in_Gamma})  to the Hamiltonian (\ref{eq:Hamiltonian_in_Fermions}) requires a  careful analysis of the boundary terms, which is given in Appendix \ref{sec:Finite N Hamiltonian}.

\subsection{Hamiltonian in k-space}
\label{subsec:fourier_transformation}

To diagonalize the quadratic fermionic Hamiltonian in (\ref{eq:Hamiltonian_in_Fermions}), we exploit the fact that the Hamiltonian is translation invariant, and go to the momentum space via defining the momentum modes $\chi^\mu_k$ as
\begin{equation}
\label{eq:Momentum_Fermion}
    \chi^\mu_a  \equiv \frac{1}{\sqrt{N}} \sum_k e^{ika} \chi^\mu_k,
\end{equation}
where the sum over $k$ runs symmetrically over both positive and negative values (see Appendix~\ref{sec:Finite N Hamiltonian} for a detailed treatment of the finite $N$ scenario). Note that $\chi^\mu_k$ are complex fermions and  satisfy the following algebra:
\begin{eqnarray}
\label{eq:Momentum_Fermion_Algebra}
    {\chi^\mu_k}^\dagger  &=& \chi^\mu_{-k}, \\
    \{ {\chi^\mu_k}^\dagger , {\chi^\nu_{k'}}\} &=& 2 \delta^{\mu \nu} \delta_{k,k'}.
\end{eqnarray}
Under this transformation, the fermionic Hamiltonian of the g-XYG  models (\ref{eq:Hamiltonian_in_Fermions}) becomes (see Appendix~\ref{sec:Finite N Hamiltonian} for the detailed calculation)
\begin{eqnarray} 
\label{eq:Hamiltonian_in_Fermions_Momentum}
H_G &=&\sum_{k>0} [\{  \text{i} \mathscr{A}_{\mu \nu} \cos k + \mathscr{S}_{\mu\nu} \sin k \} \chi^\mu_{-k} \chi^\nu_k +  \text{i}  h_i \{ \chi^{2i-1}_{-k} \chi^{2i}_k -  \chi^{2i}_{-k} \chi^{2i-1}_k \} ]. 
\end{eqnarray}
The Hamiltonian in (\ref{eq:Hamiltonian_in_Fermions_Momentum}) can also be written as 
\begin{eqnarray}
H_G=V^\dagger \pmb H V,\;\qquad \qquad   \mbox{ with } \;  V \equiv \begin{pmatrix}
        {\chi^{1}_k}   \\\vdots \\{\chi^{2d}_k} 
    \end{pmatrix}, k>0
\end{eqnarray}
where $\pmb H$ is a $2d \times 2d$ matrix given by 
\footnotesize
\begin{eqnarray}
\pmb H = \begin{pmatrix}
  \mathscr{S}_{11} \sin k & \text{i} (h_1 +  \mathscr{A}_{12} \cos k)  & \text{i} \mathscr{A}_{13} \cos k + \mathscr{S}_{13} \sin k & \dots  \\
  -\text{i} (h_1 +  \mathscr{A}_{12} \cos k) \cos k  &  \mathscr{S}_{22} \sin k &  \text{i} \mathscr{A}_{23} \cos k 
  +  \mathscr{S}_{23} \sin k & \dots \\
  - \text{i} \mathscr{A}_{13} \cos k + \mathscr{S}_{13} \sin k & - \text{i} \mathscr{A}_{32} \cos k + \mathscr{S}_{23} \sin k & \mathscr{S}_{33} \sin k & \dots\\
  \vdots & \vdots & \vdots 
  \end{pmatrix}. 
\label{eq:matrix_to_be_diagonalized}  
\end{eqnarray}
\normalsize
As mentioned in Appendix \ref{sec:Appedix:More general Hamiltonian}, with no loss of generality, we can choose $\mathscr{S}_{12}=\mathscr{S}_{34}=\dots=0$. Since these manipulations have rendered the problem of diagonalizing the Hamiltonian (\ref{eq:Hamiltonian_in_Gamma}) (acting on a $2^{Nd}$ dimensional Hilbert space) to simply diagonalizing the $2d \times 2d$ hermitian matrix  $\pmb H$, we term the model {\it solvable}. We will explicitly solve for the $d=2$ case in the next section. We mention here that all of this can be repeated with complex fermions instead of Majorana fermions and some details are given in Appendix~\ref{sec:complex_fermions}. 

\subsection{Symmetries}
\label{subsec:symmetries}

The Hamiltonian in (\ref{eq:Hamiltonian_in_Gamma}) has no symmetries for general values of the couplings $J_{\mu,\nu},h_i$, other than the discrete ${\mathbb Z}_2$ symmetry, which is given by 
\begin{eqnarray}
\Gamma^\mu_a \rightarrow - \Gamma^\mu_a. 
\end{eqnarray}
However, the Hamiltonian may enjoy certain additional symmetries for special values of these couplings. One such symmetry is the \emph{reflection symmetry}, which allows the exchange of $a\textsuperscript{th}$ site with $N-a\textsuperscript{th}$ site. For example, in the XY model of (\ref{eq:xy}), this would amount to the Hamiltonian being invariant under the transformation 
\begin{eqnarray}
\sigma^\mu_a \longleftrightarrow \sigma^\mu_{N-a},\quad  \mu = 1,2,3,\forall a. 
\label{eq:reflection_gxy}
\end{eqnarray}
This symmetry is broken once we allow for the Dzyaloshinskii–Moriya interaction term (\ref{eq:xydm}). In our model given in (\ref{eq:Hamiltonian_in_Gamma}), this symmetry can incorporated as
\begin{equation}\begin{split}
    \Gamma^{2 i-1}_a &\rightarrow -\text{i} \Gamma^{2i}_{N-a} \Gamma^{2d+1}_{N-a}, \\
    \Gamma^{2 i}_a     &\rightarrow  \text{i} \Gamma^{2i-1}_{N-a} \Gamma^{2d+1}_{N-a},
    \label{eq:reflection:gxyg}
\end{split}\end{equation}
which, for $d=1$, translates to (\ref{eq:reflection_gxy}). 
Moreover, under the transformations (\ref{eq:reflection:gxyg}), 
\begin{eqnarray}
S^i_a \rightarrow S^i_{N-a},\quad \Gamma^{2d+1}_a \rightarrow \Gamma^{2d+1}_{N-a}. 
\end{eqnarray}
The Hamiltonian in (\ref{eq:Hamiltonian_in_Gamma}) is invariant under the reflection symmetry only if the couplings satisfy
\begin{equation}\begin{split}
\label{eq:Reflection_Symmetry_constraints}
    J_{2i,2j} &= -J_{2j-1,2i-1} \\
    J_{2i-1,2j} &= J_{2j-1,2i}\\
    J_{2i,2j-1} &= J_{2j,2i-1}
\end{split}
\end{equation}
Equivalently, in terms of couplings $\mathscr{S}_{\mu \nu}, \mathscr{A}_{\mu,\nu}$ 
\begin{eqnarray}
\label{eq:Reflection_Symmetry_constraints_SA}
\mathscr{S}_{2i,2j} &=& - \mathscr{S}_{2i-1,2j-1}, \nonumber\\  \mathscr{S}_{2i,2j-1} &=& \mathscr{S}_{2i-1,2j},\nonumber \\
\mathscr{A}_{2i,2j} &=& \mathscr{A}_{2i-1,2j-1}, \nonumber\\   \mathscr{A}_{2i,2j-1} &=& - \mathscr{A}_{2i-1,2j}.
\end{eqnarray}
We will work with reflection symmetric Hamiltonians when we solve the $d=2$ model explicitly in the next section. 

\section{\lowercase{g}-XYG Models for \texorpdfstring{$d=2$}{d=2}} 
\label{sec:special_case}

In the previous section, we reduced the problem of solving for the spectrum of the Hamiltonian $H_G$ in (\ref{eq:Hamiltonian_in_Gamma}), a $2^{Nd} \times 2^{Nd}$ matrix, to solving for the eigenvalues of a $2d \times 2d$ matrix $\pmb H$ given in (\ref{eq:matrix_to_be_diagonalized}). For $d=1$, solving for the eigenvalues of the $2\times 2$ matrix can be done analytically, which results in the known spectrum of g-XY models. In this section, we focus on the more complicated case of $d=2$. 

Before we go about solving the Hamiltonian, we comment on the physical interpretation of the $d=2$ model. Recall that for $d=2$, the Hilbert space can be taken to consist of two spin half degrees of freedom (say $\sigma, \tilde{\sigma}$) per site. The most general nearest neighbour Hamiltonian that can be written on this Hilbert space is 
\begin{eqnarray}
H  = \sum_{i,j, k, l =1, a}^4 J_{i j kl} \  (\sigma^i_a \sigma^{k}_{a+1}) \otimes  (\tilde{\sigma}^j_a \tilde{\sigma}^{l}_{a+1}) + \sum_{i,j,a} h_{ij} \sigma^i_a \otimes \tilde{\sigma}^j_a
\end{eqnarray}
here $i,j,\dots$ are indices running from $1,\dots 4$ and $\sigma^0 \equiv  {\mathbb I}_{2 \times 2 }$. There are $4^4 + 4^2 = 272$ coupling constants in this Hamiltonian\footnote{The number of independent coupling constants can be reduced by rotating the Pauli matrices etc}. As mentioned before the g-XYG Hamiltonian spans a $4d^2 + d = 18$ parameter subspace of this general Hamiltonian. To get a sense of what the g-XYG interactions look like in the above conventions,  consider the $J_{12}$ term. The contribution to the g-XYG Hamiltonian (\ref{eq:Hamiltonian_in_Gamma}) after substituting the representation (\ref{eq:Gamma_with_Pauli}) is given by 
\begin{equation}
\begin{split}
H_G \supset & - \text{i} J_{12} \Gamma^{1}_a \Gamma^{5}_a \Gamma^{2}_{a+1} \\
& = -  J_{12} (\sigma^2_a \sigma^2_{a+1}) \otimes (\tilde{\sigma}^3_a)
\end{split}
\end{equation}

What we show below is that this $18$ parameter subspace is exactly solvable. {We also mention here that the 4 dimensional Hilbert space can be equivalently thought of as a spin-$3/2$ system since the Gamma matrices $\Gamma^1,\dots \Gamma^5$ can be represented by bilinear combinations of spin-$3/2$ operators - see \cite{2004PhRvB..69w5206M}}

\subsection{Hamiltonian}

For $d=2$, assuming the reflection symmetry and imposing the constraints given in (\ref{eq:Reflection_Symmetry_constraints_SA}), we obtain  \footnote{Via the choice of rotations as in Appendix~\ref{sec:Appedix:More general Hamiltonian}, we can set $\mathscr{S}_{12}=\mathscr{S}_{34}=0$, $\mathscr{S}_{22} = -\mathscr{S}_{11}, \mathscr{S}_{44} = -\mathscr{S}_{33}, \mathscr{S}_{24}= -\mathscr{S}_{13}, \mathscr{S}_{23} = \mathscr{S}_{14}, \mathscr{A}_{24} = \mathscr{A}_{13}, \mathscr{A}_{23}=-\mathscr{A}_{14}$}
\begin{footnotesize}
\begin{equation}
\pmb H  = \begin{pmatrix}
\mathscr{S}_{11} \sin k & \text{i} \left( h_1 + \mathscr{A}_{12} \cos k \right) & \text{i} \mathscr{A}_{13} \cos k +\mathscr{S}_{13} \sin k & \text{i} \mathscr{A}_{14} \cos k  +\mathscr{S}_{14} \sin k  \\
-\text{i} \left( h_1 +  A_{12} \cos k \right) & -\mathscr{S}_{11} \sin k & - \text{i} A_{14} \cos k +\mathscr{S}_{14} \sin k & \text{i} \mathscr{A}_{13} \cos k -\mathscr{S}_{13} \sin k  \\
 - \text{i} \mathscr{A}_{13} \cos k +\mathscr{S}_{13} \sin k &    \text{i} \mathscr{A}_{14} \cos k +\mathscr{S}_{14} \sin k  & \mathscr{S}_{33} \sin k & \text{i} \left( h_2 + \mathscr{A}_{34} \cos k \right) \\
 -\text{i}\mathscr{A}_{14} \cos k  +\mathscr{S}_{14} \sin k  &  -\text{i} \mathscr{A}_{13} \cos k-\mathscr{S}_{13} \sin k & - \text{i} \left( h_2 + \mathscr{A}_{34} \cos k \right)  & - \mathscr{S}_{33 } \sin k\\
\end{pmatrix}
\end{equation}
\end{footnotesize}
It is convenient to define the following quantities  
\begin{eqnarray}
\label{eq:F_k}
2F_k & = & (h_1+\mathscr{A}_{12} \cos k)^2+(h_2+\mathscr{A}_{34} \cos k )^2+2(\mathscr{A}_{13}^2+\mathscr{A}_{14}^2) \cos^2k\nonumber\\ &&+\left(\mathscr{S}_{11}^2+\mathscr{S}_{33}^2+2 \mathscr{S}_{13}^2 + 2 \mathscr{S}_{14}^2 \right)  \sin^2k,\\
\label{eq:G_k}
G_k & = &
\mbox{det } {\pmb H}
\end{eqnarray}
with $F_k \ge 0$. By computing the characteristic polynomial for $\pmb H$, one can obtain the eigenvalues to be $\{\epsilon_\pm(k) , - \epsilon_\pm(k)\}$, where
\begin{equation}
\epsilon_{\pm}(k) =\sqrt{ F_k \pm \sqrt{F_k^2 - G_k} }.
\end{equation}
Note that for
\begin{eqnarray}
G_k=0, 
\label{eq:gap_closing}
\end{eqnarray}
the eigenvalue $\epsilon_-$ which is also the energy gap between the ground and the first excited state vanishes, which corresponds to a quantum phase transition. From the eigenvectors, one can find the corresponding new quasiparticles, say, $b^\pm_k$'s and $c^\pm_k$'s for the positive energy modes and negative energy modes respectively, in terms of which one can express the Hamiltonian $H_G$ as~\cite{Zvyagin2009}
\begin{equation}
    H_G  = \sum_{k>0,s = \pm} \epsilon_s(k) \left(b^{\dagger}_{s,k} b_{s,k}  - c^\dagger_{s,k} c_{s,k}\right)
\end{equation}
Various thermodynamic properties can now be extracted in a straightforward fashion from this expression. For instance, the ground state energy of the system can be computed to be
\begin{eqnarray}
    E_g = -\sum_{k>0}\left(  \epsilon_+(k) + \epsilon_-(k) \right) = -\sqrt{2} \sum_{k>0} \sqrt{F_k + \sqrt{G_k}}.
\end{eqnarray}

\subsection{Quantum Phase Transitions}\label{sec:model:phase transition}
As mentioned before, the quantum phase transitions can be diagnosed using the gap closing condition given in (\ref{eq:gap_closing}). There may exist a number of conditions over the values of the system parameters involved in (\ref{eq:G_k}) for which this condition can be satisfied, and each of these conditions will, in principle, provide a quantum phase transition occurring in the g-XYG models for $d=2$. For the purpose of demonstration, we consider the simplest critical point of the g-XYG model, which is the analogue of \emph{order-disorder transition} in the transverse-field Ising model, which takes place at the vanishing momentum, i.e $k=0$.  The gap closing condition, $G_0=0$, can then be solved to get the critical value of the system parameter $h$ as 
\begin{eqnarray}
    h_c = \frac{-\mathscr{A}_{12}-\mu^2 \mathscr{A}_{34}\pm\sqrt{4\mu^2(\mathscr{A}_{13}^2 + \mathscr{A}_{14}^2)+(\mathscr{A}_{12}-\mu^2 \mathscr{A}_{34})^2}}{2\mu},
    \label{eq:critical_h}
\end{eqnarray}
where we have defined $\mu,h$ via $h_1 \equiv \mu h, h_2 \equiv \frac{h}{\mu}$. Note that $h_c$ is \textit{always} real, regardless of the choice of the values of the other coupling constants. 

\begin{figure*}
    \centering
    \includegraphics[width=0.7\textwidth]{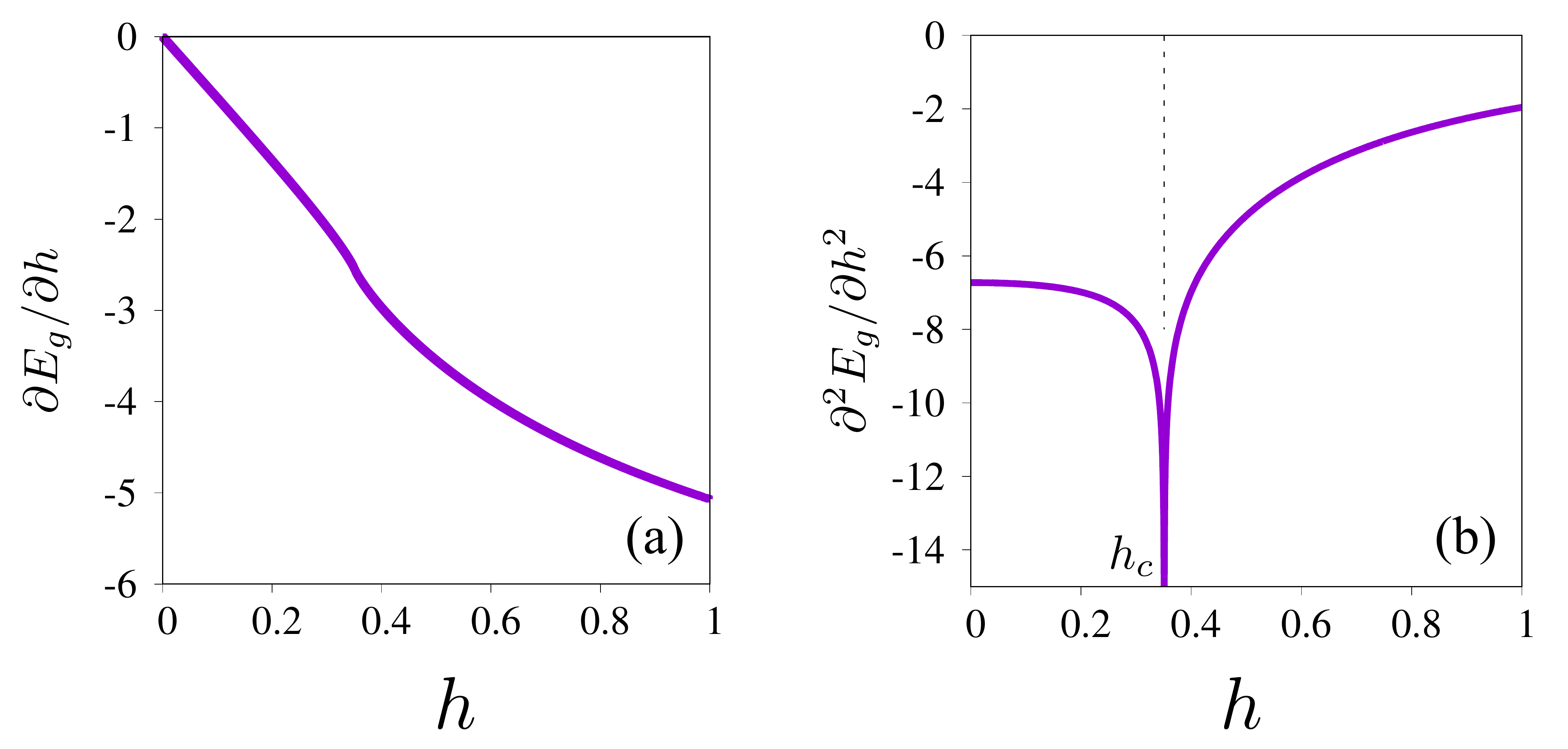}
    \caption{Variation of (a) $\partial E_g/\partial h$ and (b) $\partial^2 E_g/\partial h^2$ as functions of $h$, setting $\mu = 2$, $\mathscr{A}_{12}=1$, $\mathscr{A}_{13}=1$, $\mathscr{A}_{14}=1$, $\mathscr{A}_{34}=1$, $\mathscr{S}_{11}=1$, $\mathscr{S}_{33}=1$, $\mathscr{S}_{13}=1$, and $\mathscr{S}_{14}=1$. The quantity $\partial^2 E_g/\partial h^2$ exhibits a divergence at $h_c=0.350781$, indicating a quantum phase transition.}
    \label{fig:qpt}
\end{figure*}

Motivated by the fact that the derivatives of two-point correlation functions and single-site magnetizations provide signatures of quantum phase transitions via non-analytic behaviours,  we probe the analogous quantities in the g-XYG models. The expectation value of $\langle S_a^i\rangle$, which is the analogue of $\langle \sigma^3_a\rangle$ of the g-XY model, is obtained from  $\partial E_g/\partial h$, and is plotted as a function of $h$ in Fig.~\ref{fig:qpt}(a), where we fix $\mu=2$, and the values of the rest of the coupling constants are set to $1$, which leads to $h_c=0.350781$ (from (\ref{eq:critical_h})). At $h=h_c$, variation of $\partial E_g/\partial h$ as a function of $h$ changes from convex to concave, thereby indicating a non-analytic behaviour of $\partial^2 E_g/\partial h^2$, as shown in Fig. \ref{fig:qpt}(b). This {\it susceptibility} shows $\sim -\log|h-h_c|$  divergence near $h=h_c$.

\paragraph*{
Critical exponents}

The analysis above clearly indicates the presence of a quantum critical point
in the g-XYG model. The next natural question is to extract various
critical exponents. For this, let us analyze the behavior of the gap (=$2\epsilon_-$) in various limits. At the critical point $h=h_c$, it can be easily checked that $\epsilon_-$ vanishes linearly with $k$ when expanded 
around the critical mode $k=0$, i.e., $\epsilon_-\propto k$ upto first order in k. Clearly, this gives the value of the dynamical critical exponent z to be equal to unity. On the other hand, for the critical
mode $k=0$, the gap vanishes as $(h-h_c)$ so that $\nu z=1$, which implies that the correlation
length exponent $\nu$ is also equal to unity.

\section{Results and Conclusions}
\label{sec:results}
 
In this work, we have presented an exactly solvable generalization of the class of 1D quantum XY and Ising-like models by associating higher-dimensional Hilbert spaces to each lattice site, via replacing the Pauli matrices with Gamma matrices. Using Jordan-Wigner transformation, we have fermionized the model, and have subsequently solved it. We have illustrated an \emph{Ising-like} quantum phase transition in the model for $d=2$ with a specific set of system parameters.   

We end with a discussion on possible future directions. Within the class of models explored in this paper, the $d=2$ case provides a set of exactly solvable models with a 16-dimensional parameter space. A thorough exploration of this space may reveal more quantum phase transitions different from the one reported in this work. Besides, note that the 1D Jordan-Wigner transformations are useful in higher dimensional models too, for example,  the 2D Kitaev model on honeycomb lattice~\cite{snakelike1,snakelike2}, which can be solved via a 1D Jordan-Wigner transformations on a special path\footnote{We thank Kedar Damle for bringing this reference to our attention.}. Since $\Gamma$ matrix generalizations have been proposed also for higher dimensional models and more general lattices (see, for example, ~\cite{GammaMatrixKitaev,GammaMatrixSquare,rubylattice,kagomelattice}, and also~\cite{2009PhRvB..79u4440N} for a more general approach\footnote{We thank Vikram Tripathi for bringing this reference to our attention}), it would be interesting to explore whether the generalizations we have described using the Jordan-Wigner transformation are applicable in such higher dimensional contexts.

\acknowledgments

We acknowledge Pram Milan P Robin for collaboration in the early stages of the work. We thank Kedar Damle, Amit Dutta, R Loganayagam, Vikram Tripathi for useful discussions and insightful comments.

\appendix
\section{Fermionization of Hamiltonian} \label{sec:Finite N Hamiltonian}

In this Appendix, we give more details on the fermionization of the Hamiltonian. Though we are interested in the thermodynamic limit, we will work at finite $N$ in this Appendix. Recall that the $\Gamma$ matrices constructing the g-XYG models satisfy the periodicity condition
\begin{eqnarray}
\label{eq:Spin Periodicity}
\Gamma^\mu_{N+1} = \Gamma^\mu_N, \quad \forall \mu=1,2,\cdots,2d. 
\end{eqnarray}
For the Jordan-Wigner fermions defined in (\ref{eq:Jordan_Wigner_Fermions_from_spins}), this translates to
\begin{equation}
\label{eq:Fermion_Periodicity}
    \chi^\mu_{N+1}   = W \chi^\mu_1, 
\end{equation}
where $W \equiv \prod_{a=1}^{N}\Gamma^{2d+1}_{a}$. It is easy to see that $W^2=1$, and hence it's eigenvalues (say $w$) are $\pm 1$.  Consequently, the g-XYG model, when written in terms of fermions consists of two sectors: 
\begin{equation}\label{eq:Fermionic periodicity}
    \begin{split}
        \chi^\mu_{N+1} = - \chi_N,& \qquad \mbox{for} \qquad w  = -1, \\
        \chi^\mu_{N+1} =  \chi_N,& \qquad \mbox{for} \qquad w  = +1, \\
    \end{split}
\end{equation}
i.e., a sector with periodic fermions, and another with antiperiodic fermions. To see the sectors more clearly, let us define $H^\pm_G$ via 
\begin{eqnarray}
H_G = \frac{ 1+ W}{2} H^+_G +   \frac{ 1 - W}{2} H^-_G. 
\end{eqnarray}
The Hamiltonians $H^\pm_G$ written in terms of fermions become 
\begin{eqnarray}
\label{eq:Hamiltonian_in_Fermions_each_sectors}
H_G^\pm = -\text{i}\sum_{\mu,\nu,a}\ \text{J}_{\mu\nu} \chi^\mu_a \chi^\nu_{a+1} + \text{i}\sum_{i,a}  h_i \chi^{2i-1}_a \chi^{2i}_a.
\end{eqnarray}
Hamiltonian in each sector is thus quadratic in terms of fermions and hence simple to solve. - we just need to move to Fourier basis. However, the periodicity condition enforces that the momentum modes are either integers, or half integers. More precisely, we have\footnote{Note that in the main text, we label the  momentum modes by $k$ which was continuous whereas here the label is $q$ which are integers.}
\begin{eqnarray}
\chi^\mu_a &=& {1 \over \sqrt N} \sum_{q = -{(N-1)\over 2} }^{{(N-1)\over 2}} e^{\text{i}2\pi q a \over N} \chi^\mu_{q}, \qquad \mbox{for} \qquad w=1,\nonumber \\
\chi^\mu_a &=& {1 \over \sqrt N} \sum_{q = -{(N-2)\over 2}}^{{N\over 2}} e^{\text{i}2\pi q a \over N} \chi^\mu_{q}, \qquad \mbox{for} \qquad w=-1. 
\end{eqnarray}
Here, we have assumed that $N$ is odd, although the case of even $N$ can also be done analogously and does not change the conclusions. The modes satisfy $\chi^\mu_{q} = \chi^\mu_{q+N}$, i.e they sit on a periodic lattice. Let us denote the momentum lattice of the $w$ sector by ${\mathbb Z}^{(w)}$. The modes which satisfy $\chi^\mu_q = \chi^\mu_{-q}$ play a special role in what follows and let us denote this mode by $q_0$ - i.e $q_0 = 0$ for $w=1$ sector and $q_0 = {N \over 2}$ for $w=-1$ sector. 

It is easy to check that 
\begin{equation}\begin{split}
{\chi^\mu_q}^\dagger  &= \chi^\mu_{-q}, \\
\{ {\chi^\mu_q}^\dagger , \chi^\nu_{q'} \} & = 2\delta^{\mu \nu} \delta_{q,q'}.
\end{split}\end{equation}
Hence $\chi^\mu_q$ can be treated as complex fermions.  Note, however, that $\chi^\mu_{q_0}$ fermion is still a Majorana fermion. To solve the fermionic Hamiltonian, it is useful to note that for any $b$ (sector is denoted by $\pm$ below) ,
\begin{equation}\begin{split}
\sum_a \chi^\mu_a \chi^\nu_{a+b}
&= \sum_{q \in {\mathbb Z}^{(w)}_+ } \left( e^{2\pi \text{i} q b  \over N} \chi^\mu_{-q} \chi^\nu_q - e^{-2\pi \text{i} qb \over N} \chi^\nu_{-q} \chi^\mu_{q} \right) \\
& \qquad +{1 \over 2} \left( e^{2\pi \text{i} q_0 b \over N} \chi^\mu_{q_0} \chi^\nu_{q_0} - e^{-2\pi \text{i} q_0 b \over N} \chi^\nu_{q_0} \chi^\mu_{q_0} \right) + \kappa \delta^{\mu\nu}. 
\end{split}
\end{equation}
The constant $\kappa$ is not relevant in the discussion below, since it just shifts the Hamiltonian by a constant. Thus, the Hamiltonian given in (\ref{eq:Hamiltonian_in_Fermions_each_sectors}) becomes,
\begin{equation}
    \begin{split}
        H_G^w & =   \sum_{q \in {\mathbb Z}^{(w)}_+ } \left( \text{i} \cos{2\pi q \over N} \mathscr{A}_{\mu \nu} +\sin{2\pi q \over N}  \mathscr{S}_{\mu \nu} \right) \chi^\mu_{-q} \chi^\nu_q + {\text{i}\over 2} \cos{2\pi q_0 \over N} \mathscr{A}_{\mu \nu} \chi^\mu_{q_0} \chi^\nu_{q_0} \\
        & \qquad + \text{i}h_i \sum_{q \in {\mathbb Z}^{(w)}_+ } \left( \chi^{2i-1}_{-q} \chi^{2i}_q - \chi^{2i}_{-q} \chi^{2i-1}_{q} \right)+{\text{i} h_i \over 2}  \left( \chi^{2i-1}_{q_0} \chi^{2i}_{q_0} - \chi^{2i}_{q_0} \chi^{2i-1}_{q_0} \right).
    \end{split}
\end{equation}
Here we have used $J_{\mu \nu} = {\mathscr{S}_{\mu \nu} - \mathscr{A}_{\mu \nu}  \over 2}$. The thermodynamic limit is $N \rightarrow \infty$, with $k = {2\pi q \over N}$ kept fixed, and hence one can work with the Hamiltonian 
\begin{equation}
    H_G^w  =    \sum_{\mu, \nu,k>0}  \left( \text{i} \cos(k)  \mathscr{A}_{\mu \nu} + \sin(k)  \mathscr{S}_{\mu \nu} \right) \chi^\mu_{-k} \chi^\nu_k + \text{i} \sum_{i,k>0} h_i \left( \chi^{2i-1}_{-k} \chi^{2i}_k- \chi^{2i}_{-k} \chi^{2i-1}_{k} \right),  
\end{equation}
where we have let $\chi^\mu_q \rightarrow \chi^\mu_k$. Also, the distinction between $w=\pm 1$ sectors goes away in this limit. 

\section{Majorana to complex fermions} 
\label{sec:complex_fermions}
In this appendix, we rewrite the equations we obtained in terms of complex fermions, instead of Majorana fermions.  Let us define the complex fermions $c^{i}_a$ via
\begin{equation}
    \begin{split}
        c^{i}_a &= {\chi^{2i-1}_a  + i \chi^{2i}_a \over \sqrt 2},  \\
        {c^{i}_a }^\dagger & = {\chi^{2i-1}_a  - i \chi^{2i}_a \over \sqrt 2}.  
    \end{split}
\end{equation}
The relation between complex fermions and the matrices $\Gamma^\mu_a$ can be read off from the Jordan-Wigner transformation given in (\ref{eq:Jordan_Wigner_Fermions_from_spins}). We can also write the Fourier modes of complex fermions in terms of Fourier modes of Majorana fermions as 
\begin{equation}
    \begin{split}
        c^{i}_k &= {\chi^{2i-1}_k  + i \chi^{2i}_k \over \sqrt 2},  \\
        {c^{i}_k }^\dagger & = {\chi^{2i-1}_{-k}  - i \chi^{2i}_{-k} \over \sqrt 2}.  
    \end{split}
\end{equation}
Denoting $c^{i+}_k = {c^{i}_k}^\dagger$ and $c^{i-}_k = c^i_k$, we can rewrite the Hamiltonian in (\ref{eq:Hamiltonian_in_Fermions_Momentum}) in terms of the complex fermion modes as
\begin{equation}
\begin{split}
    H_G  &= \frac{1}{2} \sum^{d}_{i,j=1,k>0,s=\pm,\tilde{s}=\pm} \kappa^{s\tilde{s}}_{ij} c^{is}_{sk} c^{j\tilde{s}}_{-\tilde{s}k} + \sum^{d}_{i,k>0} h_{i} \{c^{i\dagger}_{k} c^{i}_{k} -c^{i}_{-k}c^{i\dagger}_{-k}\}, \\
    \end{split}
\end{equation}
where
\begin{equation}
    \begin{split}
            \kappa^{s\tilde{s}}_{ij} = \kappa_{2i-1,2j-1} + \text{i}\tilde{s}\kappa_{2i-1,2j} + \text{i}s\kappa_{2i,2j-1} - s\tilde{s}\kappa_{2i,2j}
    \end{split}
\end{equation}
with
\begin{equation}
    \begin{split}
            \kappa_{\mu\nu} = \{  \text{i} \mathscr{A}_{\mu \nu} \cos k + \mathscr{S}_{\mu\nu} \sin k \}
    \end{split}
\end{equation}

\section{A more generic Hamiltonian}\label{sec:Appedix:More general Hamiltonian}

We can actually work with the more general Hamiltonian
\begin{eqnarray}
\label{eq:Hamiltonian_in_Gamma_General}
    H_G = - \text{i}\sum_{a}\sum_{\mu,\nu}^{} \text{J}_{\mu\nu} \Gamma_{a}^{\mu} \Gamma_{a}^{2d+1} \Gamma_{a+1}^{\nu} +
    \text{i}  \sum_{a}\sum_{\mu \nu} h_{\mu \nu} \Gamma^\mu_a \Gamma^\nu_a,  
\end{eqnarray}
With no loss of generality, we can choose $h_{\mu \nu} = - h_{\nu \mu}$ since $\{\Gamma^\mu_a, \Gamma^\nu_a \}=2 \delta^{\mu \nu}$. One can easily follow through the steps given in section~\ref{sec:model} and see that this gives rise to a Hamiltonian which is quadratic in fermions. We will see below, that the number of independent coupling constants are smaller than what one expects by looking at (\ref{eq:Hamiltonian_in_Gamma_General}).

Working with a {\it rotated} set of $\Gamma$ matrices given by 
\begin{equation}
    \tilde{\Gamma}^\mu_a \equiv \sum_\nu R_{\mu \nu} \Gamma^{\nu}_a,
\end{equation}
where $R_{\mu \nu}$ are matrix elements of a real $2d \times 2d$ rotation matrix $\pmb R$ satisfying $\pmb R^T \pmb R = \pmb I$, it is easy to verify they satisfy the same algebra as in (\ref{Clifford Algebra}). The Hamiltonian (\ref{eq:Hamiltonian_in_Gamma_General}) written in terms of the rotated $\tilde{\Gamma}$ matrices still retains its form, but with the coupling constants $\tilde J_{\mu \nu}$ and $\tilde h_{\mu \nu}$. Let us denote the matrix formed by $J_{\mu \nu}$ to be $\pmb J$ and so on. It is easy to see that 
\begin{equation}
    \pmb {\tilde J} = \pmb R^T \pmb J \pmb R, \qquad \pmb {\tilde h} = \pmb R^T \pmb h \pmb R 
\end{equation}

We can always choose $\pmb R$ such that  $h_{\mu \nu}$ can be brought to a block diagonal form \cite{doi:10.1063/1.1724294}
\begin{eqnarray}\label{eq:Block Diagonal h}
\begin{pmatrix}
0 & h_1 & 0 & 0 &0 & \dots \\
- h_1 & 0 & 0 & 0  &0 & \dots  \\
0 & 0 & 0 & h_2 & 0 & \dots  \\
0 & 0 & -h_2 & 0 & 0 & \dots  \\
\vdots &\ddots &\ddots &\ddots &\ddots & \ddots 
\end{pmatrix} 
\end{eqnarray}
for some constants $h_1,h_2, \dots, h_d$. This is the form that we used in the main text (\ref{eq:Hamiltonian_in_Gamma}). However, the analysis above shows that we have not exhausted all the redefinitions yet - a further transformation by an $\pmb R$ matrix of the form 
\begin{eqnarray}
\begin{pmatrix}
\cos \theta_1 & \sin \theta_1 & 0 & 0 &0 & \dots \\
- \sin \theta_1 & \cos \theta_1 & 0 & 0  &0 & \dots  \\
0 & 0 & \cos \theta_2 & \sin \theta_2 & 0 & \dots  \\
0 & 0 & -\sin \theta_2 & \cos \theta_2 & 0 & \dots  \\
\vdots &\ddots &\ddots &\ddots &\ddots & \ddots 
\end{pmatrix}
\end{eqnarray}
keeps the form of $\pmb h$ in (\ref{eq:Block Diagonal h}) invariant. This freedom can be used to further simplify $J_{\mu \nu}$. Defining the symmetric and antisymmetric combinations of $J_{\mu \nu}$ via 
\begin{eqnarray}
\mathscr{S}_{\mu \nu} = J_{\mu \nu} + J_{\nu \mu} ,\qquad \mathscr{A}_{\mu \nu} = J_{\nu \mu} - J_{\mu \nu}  
\end{eqnarray}
One can use the above mentioned freedom to set $\mathscr{S}_{i,i+1} = 0$. This reduces the number of independent couplings to $4d^2$. 

\bibliography{ref_mod}

\begin{thebibliography}{53}%
\makeatletter
\providecommand \@ifxundefined [1]{%
 \@ifx{#1\undefined}
}%
\providecommand \@ifnum [1]{%
 \ifnum #1\expandafter \@firstoftwo
 \else \expandafter \@secondoftwo
 \fi
}%
\providecommand \@ifx [1]{%
 \ifx #1\expandafter \@firstoftwo
 \else \expandafter \@secondoftwo
 \fi
}%
\providecommand \natexlab [1]{#1}%
\providecommand \enquote  [1]{``#1''}%
\providecommand \bibnamefont  [1]{#1}%
\providecommand \bibfnamefont [1]{#1}%
\providecommand \citenamefont [1]{#1}%
\providecommand \href@noop [0]{\@secondoftwo}%
\providecommand \href [0]{\begingroup \@sanitize@url \@href}%
\providecommand \@href[1]{\@@startlink{#1}\@@href}%
\providecommand \@@href[1]{\endgroup#1\@@endlink}%
\providecommand \@sanitize@url [0]{\catcode `\\12\catcode `\$12\catcode
  `\&12\catcode `\#12\catcode `\^12\catcode `\_12\catcode `\%12\relax}%
\providecommand \@@startlink[1]{}%
\providecommand \@@endlink[0]{}%
\providecommand \url  [0]{\begingroup\@sanitize@url \@url }%
\providecommand \@url [1]{\endgroup\@href {#1}{\urlprefix }}%
\providecommand \urlprefix  [0]{URL }%
\providecommand \Eprint [0]{\href }%
\providecommand \doibase [0]{http://dx.doi.org/}%
\providecommand \selectlanguage [0]{\@gobble}%
\providecommand \bibinfo  [0]{\@secondoftwo}%
\providecommand \bibfield  [0]{\@secondoftwo}%
\providecommand \translation [1]{[#1]}%
\providecommand \BibitemOpen [0]{}%
\providecommand \bibitemStop [0]{}%
\providecommand \bibitemNoStop [0]{.\EOS\space}%
\providecommand \EOS [0]{\spacefactor3000\relax}%
\providecommand \BibitemShut  [1]{\csname bibitem#1\endcsname}%
\let\auto@bib@innerbib\@empty
%</preamble>
\bibitem [{\citenamefont {Sachdev}(2011)}]{Sachdev2011}%
  \BibitemOpen
  \bibfield  {author} {\bibinfo {author} {\bibfnamefont {S.}~\bibnamefont
  {Sachdev}},\ }\href@noop {} {\emph {\bibinfo {title} {Quantum phase
  transitions}}}\ (\bibinfo  {publisher} {Cambridge University Press,
  Cambridge},\ \bibinfo {year} {2011})\BibitemShut {NoStop}%
\bibitem [{\citenamefont {Dutta}\ \emph {et~al.}(2015)\citenamefont {Dutta},
  \citenamefont {Aeppli}, \citenamefont {Chakrabarti}, \citenamefont
  {Divakaran}, \citenamefont {Rosenbaum},\ and\ \citenamefont
  {Sen}}]{Dutta2015}%
  \BibitemOpen
  \bibfield  {author} {\bibinfo {author} {\bibfnamefont {A.}~\bibnamefont
  {Dutta}}, \bibinfo {author} {\bibfnamefont {G.}~\bibnamefont {Aeppli}},
  \bibinfo {author} {\bibfnamefont {B.~K.}\ \bibnamefont {Chakrabarti}},
  \bibinfo {author} {\bibfnamefont {U.}~\bibnamefont {Divakaran}}, \bibinfo
  {author} {\bibfnamefont {T.~F.}\ \bibnamefont {Rosenbaum}}, \ and\ \bibinfo
  {author} {\bibfnamefont {D.}~\bibnamefont {Sen}},\ }\href@noop {} {\emph
  {\bibinfo {title} {Quantum phase transitions in transverse field spin models:
  From statistical physics to quantum information}}}\ (\bibinfo  {publisher}
  {Cambridge University Press, Cambridge, UK},\ \bibinfo {year}
  {2015})\BibitemShut {NoStop}%
\bibitem [{\citenamefont {Takahashi}(1999)}]{Takahashi1999}%
  \BibitemOpen
  \bibfield  {author} {\bibinfo {author} {\bibfnamefont {M.}~\bibnamefont
  {Takahashi}},\ }\href@noop {} {\emph {\bibinfo {title} {Thermodynamics of
  one-dimensional solvable models}}}\ (\bibinfo  {publisher} {Cambridge
  University Press},\ \bibinfo {year} {1999})\BibitemShut {NoStop}%
\bibitem [{\citenamefont {Schollw\"{o}ck}(2011)}]{Schollwock2011}%
  \BibitemOpen
  \bibfield  {author} {\bibinfo {author} {\bibfnamefont {U.}~\bibnamefont
  {Schollw\"{o}ck}},\ }\bibfield  {title} {\enquote {\bibinfo {title} {The
  density-matrix renormalization group: A short introduction},}\ }\href
  {\doibase http://doi.org/10.1098/rsta.2010.0382} {\bibfield  {journal}
  {\bibinfo  {journal} {Phil. Trans. R. Soc. A.}\ }\textbf {\bibinfo {volume}
  {369}},\ \bibinfo {pages} {2643–2661} (\bibinfo {year} {2011})}\BibitemShut
  {NoStop}%
\bibitem [{\citenamefont {Pang}(2016)}]{Pang2016}%
  \BibitemOpen
  \bibfield  {author} {\bibinfo {author} {\bibfnamefont {T.}~\bibnamefont
  {Pang}},\ }\href {\doibase 10.1088/978-1-6817-4109-3} {\emph {\bibinfo
  {title} {An introduction to quantum \uppercase{M}onte \uppercase{C}arlo
  methods}}}\ (\bibinfo  {publisher} {Morgan \& Claypool Publishers},\ \bibinfo
  {year} {2016})\BibitemShut {NoStop}%
\bibitem [{\citenamefont {Montangero}(2018)}]{Montanegro2018}%
  \BibitemOpen
  \bibfield  {author} {\bibinfo {author} {\bibfnamefont {S.}~\bibnamefont
  {Montangero}},\ }\href {\doibase 10.1007/978-3-030-01409-4} {\emph {\bibinfo
  {title} {Introduction to tensor network methods}}}\ (\bibinfo  {publisher}
  {Springer International Publishing},\ \bibinfo {year} {2018})\BibitemShut
  {NoStop}%
\bibitem [{\citenamefont {Verstraete}\ and\ \citenamefont
  {Cirac}(2006)}]{cirac_mps_2006}%
  \BibitemOpen
  \bibfield  {author} {\bibinfo {author} {\bibfnamefont {F.}~\bibnamefont
  {Verstraete}}\ and\ \bibinfo {author} {\bibfnamefont {J.~I.}\ \bibnamefont
  {Cirac}},\ }\bibfield  {title} {\enquote {\bibinfo {title} {Matrix product
  states represent ground states faithfully},}\ }\href {\doibase
  10.1103/PhysRevB.73.094423} {\bibfield  {journal} {\bibinfo  {journal} {Phys.
  Rev. B}\ }\textbf {\bibinfo {volume} {73}},\ \bibinfo {pages} {094423}
  (\bibinfo {year} {2006})}\BibitemShut {NoStop}%
\bibitem [{\citenamefont {Verstraete}\ \emph {et~al.}(2008)\citenamefont
  {Verstraete}, \citenamefont {Murg},\ and\ \citenamefont
  {Cirac}}]{cirac_mps_2008}%
  \BibitemOpen
  \bibfield  {author} {\bibinfo {author} {\bibfnamefont {F.}~\bibnamefont
  {Verstraete}}, \bibinfo {author} {\bibfnamefont {V.}~\bibnamefont {Murg}}, \
  and\ \bibinfo {author} {\bibfnamefont {J.I.}\ \bibnamefont {Cirac}},\
  }\bibfield  {title} {\enquote {\bibinfo {title} {Matrix product states,
  projected entangled pair states, and variational renormalization group
  methods for quantum spin systems},}\ }\href {\doibase
  10.1080/14789940801912366} {\bibfield  {journal} {\bibinfo  {journal}
  {Advances in Physics}\ }\textbf {\bibinfo {volume} {57}},\ \bibinfo {pages}
  {143--224} (\bibinfo {year} {2008})}\BibitemShut {NoStop}%
\bibitem [{\citenamefont {Zvyagin}(2010)}]{Zvyagin2010}%
  \BibitemOpen
  \bibfield  {author} {\bibinfo {author} {\bibfnamefont {A.~A.}\ \bibnamefont
  {Zvyagin}},\ }\href@noop {} {\emph {\bibinfo {title} {Quantum theory of
  one-dimensional spin systems}}}\ (\bibinfo  {publisher} {Cambridge
  Scientific, Cambridge, U.K.},\ \bibinfo {year} {2010})\BibitemShut {NoStop}%
\bibitem [{\citenamefont {Zvyagin}(2005)}]{doi:10.1142/p364}%
  \BibitemOpen
  \bibfield  {author} {\bibinfo {author} {\bibfnamefont {A.~A.}\ \bibnamefont
  {Zvyagin}},\ }\href {\doibase 10.1142/p364} {\emph {\bibinfo {title} {Finite
  size effects in correlated electron models}}}\ (\bibinfo  {publisher} {World
  Scientific},\ \bibinfo {year} {2005})\BibitemShut {NoStop}%
\bibitem [{\citenamefont {Giamarchi}(2004)}]{Giamarchi:743140}%
  \BibitemOpen
  \bibfield  {author} {\bibinfo {author} {\bibfnamefont {T.}~\bibnamefont
  {Giamarchi}},\ }\href {\doibase 10.1093/acprof:oso/9780198525004.001.0001}
  {\emph {\bibinfo {title} {{Quantum physics in one dimension}}}},\
  International series of monographs on physics\ (\bibinfo  {publisher}
  {Clarendon Press},\ \bibinfo {address} {Oxford},\ \bibinfo {year}
  {2004})\BibitemShut {NoStop}%
\bibitem [{\citenamefont {Franchini}(2017)}]{Franchini:2016cxs}%
  \BibitemOpen
  \bibfield  {author} {\bibinfo {author} {\bibfnamefont {F.}~\bibnamefont
  {Franchini}},\ }\href {\doibase https://doi.org/10.1007/978-3-319-48487-7}
  {\emph {\bibinfo {title} {{An introduction to integrable techniques for
  one-dimensional quantum systems}}}},\ Lecture Notes in Physics\ (\bibinfo
  {publisher} {Springer},\ \bibinfo {year} {2017})\BibitemShut {NoStop}%
\bibitem [{\citenamefont
  {\u{S}amaj}(2013)}]{doi:10.1080/00107514.2016.1259251}%
  \BibitemOpen
  \bibfield  {author} {\bibinfo {author} {\bibfnamefont {Z.~B.~L.}\
  \bibnamefont {\u{S}amaj}},\ }\href {https://doi.org/10.1017/CBO9781139343480}
  {\emph {\bibinfo {title} {Introduction to the statistical physics of
  integrable many-body systems}}}\ (\bibinfo  {publisher} {Cambridge University
  Press},\ \bibinfo {year} {2013})\BibitemShut {NoStop}%
\bibitem [{\citenamefont {Elliott}\ \emph {et~al.}(1970)\citenamefont
  {Elliott}, \citenamefont {Pfeuty},\ and\ \citenamefont {Wood}}]{Elliot1970}%
  \BibitemOpen
  \bibfield  {author} {\bibinfo {author} {\bibfnamefont {R.~J.}\ \bibnamefont
  {Elliott}}, \bibinfo {author} {\bibfnamefont {P.}~\bibnamefont {Pfeuty}}, \
  and\ \bibinfo {author} {\bibfnamefont {C.}~\bibnamefont {Wood}},\ }\bibfield
  {title} {\enquote {\bibinfo {title} {\uppercase{I}sing model with a
  transverse field},}\ }\href {\doibase 10.1103/PhysRevLett.25.443} {\bibfield
  {journal} {\bibinfo  {journal} {Phys. Rev. Lett.}\ }\textbf {\bibinfo
  {volume} {25}},\ \bibinfo {pages} {443--446} (\bibinfo {year}
  {1970})}\BibitemShut {NoStop}%
\bibitem [{\citenamefont {Pfeuty}(1970)}]{Pfeuty1970}%
  \BibitemOpen
  \bibfield  {author} {\bibinfo {author} {\bibfnamefont {P.}~\bibnamefont
  {Pfeuty}},\ }\bibfield  {title} {\enquote {\bibinfo {title} {The
  one-dimensional \uppercase{I}sing model with a transverse field},}\ }\href
  {\doibase https://doi.org/10.1016/0003-4916(70)90270-8} {\bibfield  {journal}
  {\bibinfo  {journal} {Annals of Physics}\ }\textbf {\bibinfo {volume} {57}},\
  \bibinfo {pages} {79--90} (\bibinfo {year} {1970})}\BibitemShut {NoStop}%
\bibitem [{\citenamefont {Elliott}\ and\ \citenamefont
  {Wood}(1971)}]{Elliott1971}%
  \BibitemOpen
  \bibfield  {author} {\bibinfo {author} {\bibfnamefont {R.~J.}\ \bibnamefont
  {Elliott}}\ and\ \bibinfo {author} {\bibfnamefont {C.}~\bibnamefont {Wood}},\
  }\bibfield  {title} {\enquote {\bibinfo {title} {The \uppercase{I}sing model
  with a transverse field. \uppercase{I}. high temperature expansion},}\ }\href
  {\doibase 10.1088/0022-3719/4/15/023} {\bibfield  {journal} {\bibinfo
  {journal} {J. Phys. C: Solid State Physics}\ }\textbf {\bibinfo {volume}
  {4}},\ \bibinfo {pages} {2359--2369} (\bibinfo {year} {1971})}\BibitemShut
  {NoStop}%
\bibitem [{\citenamefont {Pfeuty}\ and\ \citenamefont
  {Elliott}(1971)}]{Pfeuty1971}%
  \BibitemOpen
  \bibfield  {author} {\bibinfo {author} {\bibfnamefont {P.}~\bibnamefont
  {Pfeuty}}\ and\ \bibinfo {author} {\bibfnamefont {R.~J.}\ \bibnamefont
  {Elliott}},\ }\bibfield  {title} {\enquote {\bibinfo {title} {The
  \uppercase{I}sing model with a transverse field. \uppercase{II}. ground state
  properties},}\ }\href {\doibase 10.1088/0022-3719/4/15/024} {\bibfield
  {journal} {\bibinfo  {journal} {J. Phys. C: Solid State Physics}\ }\textbf
  {\bibinfo {volume} {4}},\ \bibinfo {pages} {2370--2385} (\bibinfo {year}
  {1971})}\BibitemShut {NoStop}%
\bibitem [{\citenamefont {Stinchcombe}(1973{\natexlab{a}})}]{Stinchcombe1973}%
  \BibitemOpen
  \bibfield  {author} {\bibinfo {author} {\bibfnamefont {R.~B.}\ \bibnamefont
  {Stinchcombe}},\ }\bibfield  {title} {\enquote {\bibinfo {title}
  {\uppercase{I}sing model in a transverse field. \uppercase{I}. basic
  theory},}\ }\href {\doibase 10.1088/0022-3719/6/15/009} {\bibfield  {journal}
  {\bibinfo  {journal} {J. Phys. C: Solid State Physics}\ }\textbf {\bibinfo
  {volume} {6}},\ \bibinfo {pages} {2459--2483} (\bibinfo {year}
  {1973}{\natexlab{a}})}\BibitemShut {NoStop}%
\bibitem [{\citenamefont {Stinchcombe}(1973{\natexlab{b}})}]{Stinchcombe1973a}%
  \BibitemOpen
  \bibfield  {author} {\bibinfo {author} {\bibfnamefont {R.~B.}\ \bibnamefont
  {Stinchcombe}},\ }\bibfield  {title} {\enquote {\bibinfo {title}
  {\uppercase{I}sing model in a transverse field. \uppercase{II}. spectral
  functions and damping},}\ }\href {\doibase 10.1088/0022-3719/6/15/010}
  {\bibfield  {journal} {\bibinfo  {journal} {J. Phys. C: Solid State Physics}\
  }\textbf {\bibinfo {volume} {6}},\ \bibinfo {pages} {2484--2506} (\bibinfo
  {year} {1973}{\natexlab{b}})}\BibitemShut {NoStop}%
\bibitem [{\citenamefont {Stinchcombe}(1973{\natexlab{c}})}]{Stinchcombe1973b}%
  \BibitemOpen
  \bibfield  {author} {\bibinfo {author} {\bibfnamefont {R~B}\ \bibnamefont
  {Stinchcombe}},\ }\bibfield  {title} {\enquote {\bibinfo {title} {Thermal and
  magnetic properties of the transverse \uppercase{I}sing model},}\ }\href
  {\doibase 10.1088/0022-3719/6/15/011} {\bibfield  {journal} {\bibinfo
  {journal} {J. Phys. C: Solid State Physics}\ }\textbf {\bibinfo {volume}
  {6}},\ \bibinfo {pages} {2507--2524} (\bibinfo {year}
  {1973}{\natexlab{c}})}\BibitemShut {NoStop}%
\bibitem [{\citenamefont {Lieb}\ \emph {et~al.}(1961)\citenamefont {Lieb},
  \citenamefont {Schultz},\ and\ \citenamefont {Mattis}}]{Lieb1961}%
  \BibitemOpen
  \bibfield  {author} {\bibinfo {author} {\bibfnamefont {E.}~\bibnamefont
  {Lieb}}, \bibinfo {author} {\bibfnamefont {T.}~\bibnamefont {Schultz}}, \
  and\ \bibinfo {author} {\bibfnamefont {D.}~\bibnamefont {Mattis}},\
  }\bibfield  {title} {\enquote {\bibinfo {title} {Two soluble models of an
  antiferromagnetic chain},}\ }\href {\doibase
  https://doi.org/10.1016/0003-4916(61)90115-4} {\bibfield  {journal} {\bibinfo
   {journal} {Annals of Physics}\ }\textbf {\bibinfo {volume} {16}},\ \bibinfo
  {pages} {407--466} (\bibinfo {year} {1961})}\BibitemShut {NoStop}%
\bibitem [{\citenamefont {Katsura}(1962)}]{Katsura1962}%
  \BibitemOpen
  \bibfield  {author} {\bibinfo {author} {\bibfnamefont {S.}~\bibnamefont
  {Katsura}},\ }\bibfield  {title} {\enquote {\bibinfo {title} {Statistical
  mechanics of the anisotropic linear \uppercase{H}eisenberg model},}\ }\href
  {\doibase 10.1103/PhysRev.127.1508} {\bibfield  {journal} {\bibinfo
  {journal} {Phys. Rev.}\ }\textbf {\bibinfo {volume} {127}},\ \bibinfo {pages}
  {1508--1518} (\bibinfo {year} {1962})}\BibitemShut {NoStop}%
\bibitem [{\citenamefont {Barouch}\ \emph {et~al.}(1970)\citenamefont
  {Barouch}, \citenamefont {McCoy},\ and\ \citenamefont
  {Dresden}}]{Barouch1970}%
  \BibitemOpen
  \bibfield  {author} {\bibinfo {author} {\bibfnamefont {E.}~\bibnamefont
  {Barouch}}, \bibinfo {author} {\bibfnamefont {B.~M.}\ \bibnamefont {McCoy}},
  \ and\ \bibinfo {author} {\bibfnamefont {M.}~\bibnamefont {Dresden}},\
  }\bibfield  {title} {\enquote {\bibinfo {title} {Statistical mechanics of the
  \uppercase{XY} model. \uppercase{I}},}\ }\href {\doibase
  10.1103/PhysRevA.2.1075} {\bibfield  {journal} {\bibinfo  {journal} {Phys.
  Rev. A}\ }\textbf {\bibinfo {volume} {2}},\ \bibinfo {pages} {1075--1092}
  (\bibinfo {year} {1970})}\BibitemShut {NoStop}%
\bibitem [{\citenamefont {Barouch}\ and\ \citenamefont
  {McCoy}(1971{\natexlab{a}})}]{Barouch1971}%
  \BibitemOpen
  \bibfield  {author} {\bibinfo {author} {\bibfnamefont {E.}~\bibnamefont
  {Barouch}}\ and\ \bibinfo {author} {\bibfnamefont {B.~M.}\ \bibnamefont
  {McCoy}},\ }\bibfield  {title} {\enquote {\bibinfo {title} {Statistical
  mechanics of the \uppercase{XY} model. \uppercase{II}. spin-correlation
  functions},}\ }\href {\doibase 10.1103/PhysRevA.3.786} {\bibfield  {journal}
  {\bibinfo  {journal} {Phys. Rev. A}\ }\textbf {\bibinfo {volume} {3}},\
  \bibinfo {pages} {786--804} (\bibinfo {year}
  {1971}{\natexlab{a}})}\BibitemShut {NoStop}%
\bibitem [{\citenamefont {Barouch}\ and\ \citenamefont
  {McCoy}(1971{\natexlab{b}})}]{Barouch1971a}%
  \BibitemOpen
  \bibfield  {author} {\bibinfo {author} {\bibfnamefont {E.}~\bibnamefont
  {Barouch}}\ and\ \bibinfo {author} {\bibfnamefont {B.~M.}\ \bibnamefont
  {McCoy}},\ }\bibfield  {title} {\enquote {\bibinfo {title} {Statistical
  mechanics of the \uppercase{XY} model. \uppercase{III}},}\ }\href {\doibase
  10.1103/PhysRevA.3.2137} {\bibfield  {journal} {\bibinfo  {journal} {Phys.
  Rev. A}\ }\textbf {\bibinfo {volume} {3}},\ \bibinfo {pages} {2137--2140}
  (\bibinfo {year} {1971}{\natexlab{b}})}\BibitemShut {NoStop}%
\bibitem [{\citenamefont {Kogut}(1979)}]{Kogut1979}%
  \BibitemOpen
  \bibfield  {author} {\bibinfo {author} {\bibfnamefont {J.~B.}\ \bibnamefont
  {Kogut}},\ }\bibfield  {title} {\enquote {\bibinfo {title} {An introduction
  to lattice gauge theory and spin systems},}\ }\href {\doibase
  10.1103/RevModPhys.51.659} {\bibfield  {journal} {\bibinfo  {journal} {Rev.
  Mod. Phys.}\ }\textbf {\bibinfo {volume} {51}},\ \bibinfo {pages} {659--713}
  (\bibinfo {year} {1979})}\BibitemShut {NoStop}%
\bibitem [{\citenamefont {Amico}\ \emph {et~al.}(2008)\citenamefont {Amico},
  \citenamefont {Fazio}, \citenamefont {Osterloh},\ and\ \citenamefont
  {Vedral}}]{Amico2008}%
  \BibitemOpen
  \bibfield  {author} {\bibinfo {author} {\bibfnamefont {L.}~\bibnamefont
  {Amico}}, \bibinfo {author} {\bibfnamefont {R.}~\bibnamefont {Fazio}},
  \bibinfo {author} {\bibfnamefont {A.}~\bibnamefont {Osterloh}}, \ and\
  \bibinfo {author} {\bibfnamefont {V.}~\bibnamefont {Vedral}},\ }\bibfield
  {title} {\enquote {\bibinfo {title} {Entanglement in many-body systems},}\
  }\href {\doibase 10.1103/RevModPhys.80.517} {\bibfield  {journal} {\bibinfo
  {journal} {Rev. Mod. Phys.}\ }\textbf {\bibinfo {volume} {80}},\ \bibinfo
  {pages} {517--576} (\bibinfo {year} {2008})}\BibitemShut {NoStop}%
\bibitem [{\citenamefont {Youngblood}\ \emph {et~al.}(1982)\citenamefont
  {Youngblood}, \citenamefont {Aeppli}, \citenamefont {Axe},\ and\
  \citenamefont {Griffin}}]{aeppli_82}%
  \BibitemOpen
  \bibfield  {author} {\bibinfo {author} {\bibfnamefont {R.~W.}\ \bibnamefont
  {Youngblood}}, \bibinfo {author} {\bibfnamefont {G.}~\bibnamefont {Aeppli}},
  \bibinfo {author} {\bibfnamefont {J.~D.}\ \bibnamefont {Axe}}, \ and\
  \bibinfo {author} {\bibfnamefont {J.~A.}\ \bibnamefont {Griffin}},\
  }\bibfield  {title} {\enquote {\bibinfo {title} {Spin dynamics of a model
  singlet ground-state system},}\ }\href {\doibase 10.1103/PhysRevLett.49.1724}
  {\bibfield  {journal} {\bibinfo  {journal} {Phys. Rev. Lett.}\ }\textbf
  {\bibinfo {volume} {49}},\ \bibinfo {pages} {1724--1727} (\bibinfo {year}
  {1982})}\BibitemShut {NoStop}%
\bibitem [{\citenamefont {Porras}\ and\ \citenamefont
  {Cirac}(2004)}]{Porras2004}%
  \BibitemOpen
  \bibfield  {author} {\bibinfo {author} {\bibfnamefont {D.}~\bibnamefont
  {Porras}}\ and\ \bibinfo {author} {\bibfnamefont {J.~I.}\ \bibnamefont
  {Cirac}},\ }\bibfield  {title} {\enquote {\bibinfo {title} {Effective quantum
  spin systems with trapped ions},}\ }\href {\doibase
  10.1103/PhysRevLett.92.207901} {\bibfield  {journal} {\bibinfo  {journal}
  {Phys. Rev. Lett.}\ }\textbf {\bibinfo {volume} {92}},\ \bibinfo {pages}
  {207901} (\bibinfo {year} {2004})}\BibitemShut {NoStop}%
\bibitem [{\citenamefont {Deng}\ \emph {et~al.}(2005)\citenamefont {Deng},
  \citenamefont {Porras},\ and\ \citenamefont {Cirac}}]{Deng2005}%
  \BibitemOpen
  \bibfield  {author} {\bibinfo {author} {\bibfnamefont {X.-L.}\ \bibnamefont
  {Deng}}, \bibinfo {author} {\bibfnamefont {D.}~\bibnamefont {Porras}}, \ and\
  \bibinfo {author} {\bibfnamefont {J.~I.}\ \bibnamefont {Cirac}},\ }\bibfield
  {title} {\enquote {\bibinfo {title} {Effective spin quantum phases in systems
  of trapped ions},}\ }\href {\doibase 10.1103/PhysRevA.72.063407} {\bibfield
  {journal} {\bibinfo  {journal} {Phys. Rev. A}\ }\textbf {\bibinfo {volume}
  {72}},\ \bibinfo {pages} {063407} (\bibinfo {year} {2005})}\BibitemShut
  {NoStop}%
\bibitem [{\citenamefont {Zhang}\ \emph {et~al.}(2012)\citenamefont {Zhang},
  \citenamefont {Yung}, \citenamefont {Laflamme}, \citenamefont
  {Aspuru-Guzik},\ and\ \citenamefont {Baugh}}]{Zhang2012}%
  \BibitemOpen
  \bibfield  {author} {\bibinfo {author} {\bibfnamefont {J.}~\bibnamefont
  {Zhang}}, \bibinfo {author} {\bibfnamefont {M.-H.}\ \bibnamefont {Yung}},
  \bibinfo {author} {\bibfnamefont {R.}~\bibnamefont {Laflamme}}, \bibinfo
  {author} {\bibfnamefont {A.}~\bibnamefont {Aspuru-Guzik}}, \ and\ \bibinfo
  {author} {\bibfnamefont {J.}~\bibnamefont {Baugh}},\ }\bibfield  {title}
  {\enquote {\bibinfo {title} {Digital quantum simulation of the statistical
  mechanics of a frustrated magnet},}\ }\href {\doibase 10.1038/ncomms1860}
  {\bibfield  {journal} {\bibinfo  {journal} {Nature Communications}\ }\textbf
  {\bibinfo {volume} {3}},\ \bibinfo {pages} {880} (\bibinfo {year}
  {2012})}\BibitemShut {NoStop}%
\bibitem [{\citenamefont {Schechter}\ and\ \citenamefont
  {Stamp}(2008)}]{Schechter2008}%
  \BibitemOpen
  \bibfield  {author} {\bibinfo {author} {\bibfnamefont {M.}~\bibnamefont
  {Schechter}}\ and\ \bibinfo {author} {\bibfnamefont {P.~C.~E.}\ \bibnamefont
  {Stamp}},\ }\bibfield  {title} {\enquote {\bibinfo {title} {Derivation of the
  low-\uppercase{\emph{t}} phase diagram of
  \uppercase{L}i\uppercase{H}o$_x$\uppercase{Y}$_{1-x}$\uppercase{F}$_4$: A
  dipolar quantum \uppercase{I}sing magnet},}\ }\href {\doibase
  10.1103/PhysRevB.78.054438} {\bibfield  {journal} {\bibinfo  {journal} {Phys.
  Rev. B}\ }\textbf {\bibinfo {volume} {78}},\ \bibinfo {pages} {054438}
  (\bibinfo {year} {2008})}\BibitemShut {NoStop}%
\bibitem [{\citenamefont {Duan}\ \emph {et~al.}(2003)\citenamefont {Duan},
  \citenamefont {Demler},\ and\ \citenamefont {Lukin}}]{Duan2003}%
  \BibitemOpen
  \bibfield  {author} {\bibinfo {author} {\bibfnamefont {L.-M.}\ \bibnamefont
  {Duan}}, \bibinfo {author} {\bibfnamefont {E.}~\bibnamefont {Demler}}, \ and\
  \bibinfo {author} {\bibfnamefont {M.~D.}\ \bibnamefont {Lukin}},\ }\bibfield
  {title} {\enquote {\bibinfo {title} {Controlling spin exchange interactions
  of ultracold atoms in optical lattices},}\ }\href {\doibase
  10.1103/PhysRevLett.91.090402} {\bibfield  {journal} {\bibinfo  {journal}
  {Phys. Rev. Lett.}\ }\textbf {\bibinfo {volume} {91}},\ \bibinfo {pages}
  {090402} (\bibinfo {year} {2003})}\BibitemShut {NoStop}%
\bibitem [{\citenamefont {Simon}\ \emph {et~al.}(2011)\citenamefont {Simon},
  \citenamefont {Bakr}, \citenamefont {Ma}, \citenamefont {Tai}, \citenamefont
  {Preiss},\ and\ \citenamefont {Greiner}}]{Simon2011}%
  \BibitemOpen
  \bibfield  {author} {\bibinfo {author} {\bibfnamefont {J.}~\bibnamefont
  {Simon}}, \bibinfo {author} {\bibfnamefont {W.~S.}\ \bibnamefont {Bakr}},
  \bibinfo {author} {\bibfnamefont {R.}~\bibnamefont {Ma}}, \bibinfo {author}
  {\bibfnamefont {M.~E.}\ \bibnamefont {Tai}}, \bibinfo {author} {\bibfnamefont
  {P.~M.}\ \bibnamefont {Preiss}}, \ and\ \bibinfo {author} {\bibfnamefont
  {M.}~\bibnamefont {Greiner}},\ }\bibfield  {title} {\enquote {\bibinfo
  {title} {Quantum simulation of antiferromagnetic spin chains in an optical
  lattice},}\ }\href {\doibase 10.1038/nature09994} {\bibfield  {journal}
  {\bibinfo  {journal} {Nature}\ }\textbf {\bibinfo {volume} {472}},\ \bibinfo
  {pages} {307--312} (\bibinfo {year} {2011})}\BibitemShut {NoStop}%
\bibitem [{\citenamefont {Liao}\ \emph {et~al.}(2021)\citenamefont {Liao},
  \citenamefont {Xiong},\ and\ \citenamefont {Chen}}]{Liao2021}%
  \BibitemOpen
  \bibfield  {author} {\bibinfo {author} {\bibfnamefont {R.}~\bibnamefont
  {Liao}}, \bibinfo {author} {\bibfnamefont {F.}~\bibnamefont {Xiong}}, \ and\
  \bibinfo {author} {\bibfnamefont {X.}~\bibnamefont {Chen}},\ }\bibfield
  {title} {\enquote {\bibinfo {title} {Simulating an exact one-dimensional
  transverse \uppercase{I}sing model in an optical lattice},}\ }\href {\doibase
  10.1103/PhysRevA.103.043312} {\bibfield  {journal} {\bibinfo  {journal}
  {Phys. Rev. A}\ }\textbf {\bibinfo {volume} {103}},\ \bibinfo {pages}
  {043312} (\bibinfo {year} {2021})}\BibitemShut {NoStop}%
\bibitem [{\citenamefont {Siskens}\ \emph {et~al.}(1975)\citenamefont
  {Siskens}, \citenamefont {Capel},\ and\ \citenamefont
  {Gaemers}}]{SISKENS1975259}%
  \BibitemOpen
  \bibfield  {author} {\bibinfo {author} {\bibfnamefont {T.~J.}\ \bibnamefont
  {Siskens}}, \bibinfo {author} {\bibfnamefont {H.~W.}\ \bibnamefont {Capel}},
  \ and\ \bibinfo {author} {\bibfnamefont {K.~J.~F.}\ \bibnamefont {Gaemers}},\
  }\bibfield  {title} {\enquote {\bibinfo {title} {On a soluble model of an
  antiferromagnetic chain with \uppercase{D}zyaloshinsky interactions.
  \uppercase{I}},}\ }\href {\doibase
  https://doi.org/10.1016/0378-4371(75)90029-1} {\bibfield  {journal} {\bibinfo
   {journal} {Physica A: Statistical Mechanics and its Applications}\ }\textbf
  {\bibinfo {volume} {79}},\ \bibinfo {pages} {259--295} (\bibinfo {year}
  {1975})}\BibitemShut {NoStop}%
\bibitem [{\citenamefont {Divakaran}\ \emph {et~al.}(2008)\citenamefont
  {Divakaran}, \citenamefont {Dutta},\ and\ \citenamefont
  {Sen}}]{divakaran2008}%
  \BibitemOpen
  \bibfield  {author} {\bibinfo {author} {\bibfnamefont {U.}~\bibnamefont
  {Divakaran}}, \bibinfo {author} {\bibfnamefont {A.}~\bibnamefont {Dutta}}, \
  and\ \bibinfo {author} {\bibfnamefont {D.}~\bibnamefont {Sen}},\ }\bibfield
  {title} {\enquote {\bibinfo {title} {Quenching along a gapless line: A
  different exponent for defect density},}\ }\href {\doibase
  10.1103/PhysRevB.78.144301} {\bibfield  {journal} {\bibinfo  {journal} {Phys.
  Rev. B}\ }\textbf {\bibinfo {volume} {78}},\ \bibinfo {pages} {144301}
  (\bibinfo {year} {2008})}\BibitemShut {NoStop}%
\bibitem [{\citenamefont {Kopp}\ and\ \citenamefont
  {Chakravarty}(2005)}]{Kopp2005}%
  \BibitemOpen
  \bibfield  {author} {\bibinfo {author} {\bibfnamefont {A.}~\bibnamefont
  {Kopp}}\ and\ \bibinfo {author} {\bibfnamefont {S.}~\bibnamefont
  {Chakravarty}},\ }\bibfield  {title} {\enquote {\bibinfo {title} {Criticality
  in correlated quantum matter},}\ }\href {\doibase 10.1038/nphys105}
  {\bibfield  {journal} {\bibinfo  {journal} {Nature Physics}\ }\textbf
  {\bibinfo {volume} {1}},\ \bibinfo {pages} {53--56} (\bibinfo {year}
  {2005})}\BibitemShut {NoStop}%
\bibitem [{\citenamefont {Zvyagin}\ and\ \citenamefont
  {Skorobaga\'{t}ko}(2006)}]{Zvyagin2006}%
  \BibitemOpen
  \bibfield  {author} {\bibinfo {author} {\bibfnamefont {A.~A.}\ \bibnamefont
  {Zvyagin}}\ and\ \bibinfo {author} {\bibfnamefont {G.~A.}\ \bibnamefont
  {Skorobaga\'{t}ko}},\ }\bibfield  {title} {\enquote {\bibinfo {title}
  {Exactly solvable quantum spin model with alternating and multiple spin
  exchange interactions},}\ }\href {\doibase 10.1103/PhysRevB.73.024427}
  {\bibfield  {journal} {\bibinfo  {journal} {Phys. Rev. B}\ }\textbf {\bibinfo
  {volume} {73}},\ \bibinfo {pages} {024427} (\bibinfo {year}
  {2006})}\BibitemShut {NoStop}%
\bibitem [{\citenamefont {Zvyagin}(2009)}]{Zvyagin2009}%
  \BibitemOpen
  \bibfield  {author} {\bibinfo {author} {\bibfnamefont {A.~A.}\ \bibnamefont
  {Zvyagin}},\ }\bibfield  {title} {\enquote {\bibinfo {title} {Quantum phase
  transitions in an exactly solvable quantum-spin biaxial model with multiple
  spin interactions},}\ }\href {\doibase 10.1103/PhysRevB.80.014414} {\bibfield
   {journal} {\bibinfo  {journal} {Phys. Rev. B}\ }\textbf {\bibinfo {volume}
  {80}},\ \bibinfo {pages} {014414} (\bibinfo {year} {2009})}\BibitemShut
  {NoStop}%
\bibitem [{\citenamefont {Wigner}\ and\ \citenamefont
  {Jordan}(1928)}]{Wigner1928}%
  \BibitemOpen
  \bibfield  {author} {\bibinfo {author} {\bibfnamefont {E.~P.}\ \bibnamefont
  {Wigner}}\ and\ \bibinfo {author} {\bibfnamefont {P.}~\bibnamefont
  {Jordan}},\ }\bibfield  {title} {\enquote {\bibinfo {title} {\"{U}ber das
  {P}aulische \u{A}quivalenzverbot},}\ }\href@noop {} {\bibfield  {journal}
  {\bibinfo  {journal} {Zeitschrift f\"{u}r Physik}\ }\textbf {\bibinfo
  {volume} {5}},\ \bibinfo {pages} {11} (\bibinfo {year} {1928})}\BibitemShut
  {NoStop}%
\bibitem [{\citenamefont {{Dargis}}\ and\ \citenamefont
  {{Maassarani}}(1998)}]{1998NuPhB.535..681D}%
  \BibitemOpen
  \bibfield  {author} {\bibinfo {author} {\bibfnamefont {P.}~\bibnamefont
  {{Dargis}}}\ and\ \bibinfo {author} {\bibfnamefont {Z.}~\bibnamefont
  {{Maassarani}}},\ }\bibfield  {title} {\enquote {\bibinfo {title}
  {{Fermionization and {H}ubbard models}},}\ }\href {\doibase
  10.1016/S0550-3213(98)00650-6} {\bibfield  {journal} {\bibinfo  {journal}
  {Nuclear Physics B}\ }\textbf {\bibinfo {volume} {535}},\ \bibinfo {pages}
  {681--708} (\bibinfo {year} {1998})},\ \Eprint
  {http://arxiv.org/abs/cond-mat/9806208} {arXiv:cond-mat/9806208 [cond-mat]}
  \BibitemShut {NoStop}%
\bibitem [{\citenamefont {{Bochniak}}\ \emph {et~al.}(2021)\citenamefont
  {{Bochniak}}, \citenamefont {{Ruba}},\ and\ \citenamefont
  {{Wosiek}}}]{2021arXiv210706335B}%
  \BibitemOpen
  \bibfield  {author} {\bibinfo {author} {\bibfnamefont {A.}~\bibnamefont
  {{Bochniak}}}, \bibinfo {author} {\bibfnamefont {B.}~\bibnamefont {{Ruba}}},
  \ and\ \bibinfo {author} {\bibfnamefont {J.}~\bibnamefont {{Wosiek}}},\
  }\bibfield  {title} {\enquote {\bibinfo {title} {{Bosonization of
  \uppercase{M}ajorana modes and edge states}},}\ }\href
  {https://arxiv.org/abs/2107.06335} {\bibfield  {journal} {\bibinfo  {journal}
  {arXiv:2107.06335}\ } (\bibinfo {year} {2021})}\BibitemShut {NoStop}%
\bibitem [{\citenamefont {Zhu}(2016)}]{bdg2016}%
  \BibitemOpen
  \bibfield  {author} {\bibinfo {author} {\bibfnamefont {Jian-Xin}\
  \bibnamefont {Zhu}},\ }\href@noop {} {\emph {\bibinfo {title}
  {\uppercase{B}ogoliubov-de \uppercase{G}ennes Method and Its Applications}}}\
  (\bibinfo  {publisher} {Springer},\ \bibinfo {year} {2016})\BibitemShut
  {NoStop}%
\bibitem [{\citenamefont {{Murakami}}\ \emph {et~al.}(2004)\citenamefont
  {{Murakami}}, \citenamefont {{Nagosa}},\ and\ \citenamefont
  {{Zhang}}}]{2004PhRvB..69w5206M}%
  \BibitemOpen
  \bibfield  {author} {\bibinfo {author} {\bibfnamefont {S.}~\bibnamefont
  {{Murakami}}}, \bibinfo {author} {\bibfnamefont {N.}~\bibnamefont
  {{Nagosa}}}, \ and\ \bibinfo {author} {\bibfnamefont {S.-C.}\ \bibnamefont
  {{Zhang}}},\ }\bibfield  {title} {\enquote {\bibinfo {title}
  {{\uppercase{SU}(2) non-Abelian holonomy and dissipationless spin current in
  semiconductors}},}\ }\href {\doibase 10.1103/PhysRevB.69.235206} {\bibfield
  {journal} {\bibinfo  {journal} {\prb}\ }\textbf {\bibinfo {volume} {69}},\
  \bibinfo {eid} {235206} (\bibinfo {year} {2004})}\BibitemShut {NoStop}%
\bibitem [{\citenamefont {{Feng}}\ \emph {et~al.}(2007)\citenamefont {{Feng}},
  \citenamefont {{Zhang}},\ and\ \citenamefont {{Xiang}}}]{snakelike1}%
  \BibitemOpen
  \bibfield  {author} {\bibinfo {author} {\bibfnamefont {X.-Y.}\ \bibnamefont
  {{Feng}}}, \bibinfo {author} {\bibfnamefont {G.-M.}\ \bibnamefont {{Zhang}}},
  \ and\ \bibinfo {author} {\bibfnamefont {T.}~\bibnamefont {{Xiang}}},\
  }\bibfield  {title} {\enquote {\bibinfo {title} {{Topological
  characterization of quantum phase transitions in a spin-1/2 model}},}\ }\href
  {\doibase 10.1103/PhysRevLett.98.087204} {\bibfield  {journal} {\bibinfo
  {journal} {\prl}\ }\textbf {\bibinfo {volume} {98}},\ \bibinfo {eid} {087204}
  (\bibinfo {year} {2007})}\BibitemShut {NoStop}%
\bibitem [{\citenamefont {{Chen}}\ and\ \citenamefont
  {{Hu}}(2007)}]{snakelike2}%
  \BibitemOpen
  \bibfield  {author} {\bibinfo {author} {\bibfnamefont {H.-D.}\ \bibnamefont
  {{Chen}}}\ and\ \bibinfo {author} {\bibfnamefont {J.}~\bibnamefont {{Hu}}},\
  }\bibfield  {title} {\enquote {\bibinfo {title} {{Exact mapping between
  classical and topological orders in two-dimensional spin systems}},}\ }\href
  {\doibase 10.1103/PhysRevB.76.193101} {\bibfield  {journal} {\bibinfo
  {journal} {\prb}\ }\textbf {\bibinfo {volume} {76}},\ \bibinfo {eid} {193101}
  (\bibinfo {year} {2007})}\BibitemShut {NoStop}%
\bibitem [{\citenamefont {Wu}\ \emph {et~al.}(2009)\citenamefont {Wu},
  \citenamefont {Arovas},\ and\ \citenamefont {Hung}}]{GammaMatrixKitaev}%
  \BibitemOpen
  \bibfield  {author} {\bibinfo {author} {\bibfnamefont {C.}~\bibnamefont
  {Wu}}, \bibinfo {author} {\bibfnamefont {D.}~\bibnamefont {Arovas}}, \ and\
  \bibinfo {author} {\bibfnamefont {H.-H.}\ \bibnamefont {Hung}},\ }\bibfield
  {title} {\enquote {\bibinfo {title} {$\ensuremath{\Gamma}$-matrix
  generalization of the \uppercase{K}itaev model},}\ }\href {\doibase
  10.1103/PhysRevB.79.134427} {\bibfield  {journal} {\bibinfo  {journal} {Phys.
  Rev. B}\ }\textbf {\bibinfo {volume} {79}},\ \bibinfo {pages} {134427}
  (\bibinfo {year} {2009})}\BibitemShut {NoStop}%
\bibitem [{\citenamefont {Yao}\ \emph {et~al.}(2009)\citenamefont {Yao},
  \citenamefont {Zhang},\ and\ \citenamefont {Kivelson}}]{GammaMatrixSquare}%
  \BibitemOpen
  \bibfield  {author} {\bibinfo {author} {\bibfnamefont {H.}~\bibnamefont
  {Yao}}, \bibinfo {author} {\bibfnamefont {S.-C.}\ \bibnamefont {Zhang}}, \
  and\ \bibinfo {author} {\bibfnamefont {S.~A.}\ \bibnamefont {Kivelson}},\
  }\bibfield  {title} {\enquote {\bibinfo {title} {Algebraic spin liquid in an
  exactly solvable spin model},}\ }\href {\doibase
  10.1103/PhysRevLett.102.217202} {\bibfield  {journal} {\bibinfo  {journal}
  {Phys. Rev. Lett.}\ }\textbf {\bibinfo {volume} {102}},\ \bibinfo {pages}
  {217202} (\bibinfo {year} {2009})}\BibitemShut {NoStop}%
\bibitem [{\citenamefont {{Whitsitt}}\ \emph {et~al.}(2012)\citenamefont
  {{Whitsitt}}, \citenamefont {{Chua}},\ and\ \citenamefont
  {{Fiete}}}]{rubylattice}%
  \BibitemOpen
  \bibfield  {author} {\bibinfo {author} {\bibfnamefont {S.}~\bibnamefont
  {{Whitsitt}}}, \bibinfo {author} {\bibfnamefont {V.}~\bibnamefont {{Chua}}},
  \ and\ \bibinfo {author} {\bibfnamefont {G.~A.}\ \bibnamefont {{Fiete}}},\
  }\bibfield  {title} {\enquote {\bibinfo {title} {Exact chiral spin liquids
  and mean-field perturbations of gamma matrix models on the ruby lattice},}\
  }\href {\doibase 10.1088/1367-2630/14/11/115029} {\bibfield  {journal}
  {\bibinfo  {journal} {New Journal of Physics}\ }\textbf {\bibinfo {volume}
  {14}},\ \bibinfo {eid} {115029} (\bibinfo {year} {2012})}\BibitemShut
  {NoStop}%
\bibitem [{\citenamefont {{Chua}}\ \emph {et~al.}(2011)\citenamefont {{Chua}},
  \citenamefont {{Yao}},\ and\ \citenamefont {{Fiete}}}]{kagomelattice}%
  \BibitemOpen
  \bibfield  {author} {\bibinfo {author} {\bibfnamefont {V.}~\bibnamefont
  {{Chua}}}, \bibinfo {author} {\bibfnamefont {H.}~\bibnamefont {{Yao}}}, \
  and\ \bibinfo {author} {\bibfnamefont {G.~A.}\ \bibnamefont {{Fiete}}},\
  }\bibfield  {title} {\enquote {\bibinfo {title} {{Exact chiral spin liquid
  with stable spin \uppercase{F}ermi surface on the kagome lattice}},}\ }\href
  {\doibase 10.1103/PhysRevB.83.180412} {\bibfield  {journal} {\bibinfo
  {journal} {\prb}\ }\textbf {\bibinfo {volume} {83}},\ \bibinfo {eid} {180412}
  (\bibinfo {year} {2011})}\BibitemShut {NoStop}%
\bibitem [{\citenamefont {{Nussinov}}\ and\ \citenamefont
  {{Ortiz}}(2009)}]{2009PhRvB..79u4440N}%
  \BibitemOpen
  \bibfield  {author} {\bibinfo {author} {\bibfnamefont {Z.}~\bibnamefont
  {{Nussinov}}}\ and\ \bibinfo {author} {\bibfnamefont {G.}~\bibnamefont
  {{Ortiz}}},\ }\bibfield  {title} {\enquote {\bibinfo {title} {{Bond algebras
  and exact solvability of \uppercase{H}amiltonians: Spin S=1/2 multilayer
  systems}},}\ }\href {\doibase 10.1103/PhysRevB.79.214440} {\bibfield
  {journal} {\bibinfo  {journal} {\prb}\ }\textbf {\bibinfo {volume} {79}},\
  \bibinfo {eid} {214440} (\bibinfo {year} {2009})}\BibitemShut {NoStop}%
\bibitem [{\citenamefont {Zumino}(1962)}]{doi:10.1063/1.1724294}%
  \BibitemOpen
  \bibfield  {author} {\bibinfo {author} {\bibfnamefont {B.}~\bibnamefont
  {Zumino}},\ }\bibfield  {title} {\enquote {\bibinfo {title} {Normal forms of
  complex matrices},}\ }\href {\doibase 10.1063/1.1724294} {\bibfield
  {journal} {\bibinfo  {journal} {Journal of Mathematical Physics}\ }\textbf
  {\bibinfo {volume} {3}},\ \bibinfo {pages} {1055--1057} (\bibinfo {year}
  {1962})}\BibitemShut {NoStop}%
\end{thebibliography}%

\end{document}